\documentclass[english]{IEEEtran}
\usepackage[T1]{fontenc}
\usepackage[latin9]{inputenc}
\usepackage{color}
\usepackage{array}
\usepackage{booktabs}
\usepackage{textcomp}
\usepackage{multirow}
 \usepackage[noadjust]{cite}
    
\usepackage{threeparttable}

\usepackage[caption=false]{subfig}
\usepackage{amsmath}
\usepackage{amsthm}
\usepackage{amssymb}
\usepackage[export]{adjustbox}
\usepackage{verbatim}
\usepackage{graphicx}
\usepackage{caption}
\usepackage[shortlabels]{enumitem}
\def\infinity{\rotatebox{90}{8}}
\usepackage{tikz}
\tikzstyle{io} = [fill=black,inner sep=2pt,circle]
\usetikzlibrary{positioning,calc,arrows.meta}
\usetikzlibrary{arrows,positioning}

\everymath{\displaystyle}
\usetikzlibrary{backgrounds}

\usepackage{pifont}% http://ctan.org/pkg/pifont
\newcommand{\cmark}{\ding{51}}%
\newcommand{\xmark}{\ding{55}}%

\usepackage{mathtools}
\DeclarePairedDelimiter{\abs}{\lvert}{\rvert}

\makeatletter
\def\endthebibliography{%
	\def\@noitemerr{\@latex@warning{Empty `thebibliography' environment}}%
	\endlist
}

\makeatletter
\newcommand*\bigcdot{\mathpalette\bigcdot@{.5}}
\newcommand*\bigcdot@[2]{\mathbin{\vcenter{\hbox{\scalebox{#2}{$\m@th#1\bullet$}}}}}
\makeatother

\makeatletter

\theoremstyle{plain}

\usetikzlibrary{shapes,arrows}
\tikzstyle{line}=[draw] % here
\usepackage{algorithmic}
\usepackage{amsmath,varwidth,array,ragged2e}
\usepackage{graphicx}
\usepackage{enumitem}
\usepackage{amsfonts}
\usepackage{mathrsfs}
\usepackage{url}
\usepackage{soul}
\usepackage{xcolor}
\@ifundefined{showcaptionsetup}{}{%
	\PassOptionsToPackage{caption=false}{subfig}}
\usepackage{subfig}
\makeatother
\usepackage{babel}
\providecommand{\theoremname}{Theorem}

\begin{document}

	\title{Detecting Sybil Attacks using Proofs of Work and Location in VANETs}
	
	\author{Mohamed~Baza\IEEEauthorrefmark{1},~Mahmoud~Nabil\IEEEauthorrefmark{1},~Niclas~Bewermeier\IEEEauthorrefmark{1},~Kemal Fidan\IEEEauthorrefmark{2},\\~Mohamed Mahmoud\IEEEauthorrefmark{1},~Mohamed~Abdallah\IEEEauthorrefmark{3}
		\\
			
		\IEEEauthorblockA{\IEEEauthorrefmark{1}Department of Electrical and Computer Engineering, Tennessee Tech University, Cookeville, TN, USA
		}\\

		\IEEEauthorblockA{\IEEEauthorrefmark{2}Department of Electrical Engineering and Computer Science, University of Tennessee, Knoxville, TN, USA}\\
		
		\IEEEauthorblockA{\IEEEauthorrefmark{3}Division of Information and Computing Technology, College of Science and Engineering, HBKU, Doha, Qata}\\
	 \vspace{-8mm}}
	
	\maketitle
	\begin{abstract}

%However, a malicious vehicle may launch a Sybil attack by masquerading to be multiple simultaneous vehicles. If vehicles or RSUs are unable to detect such an attack, this will lead to reporting many forged number of vehicles in the area in which Sybil attack occur which may result in vital consequences such as severe car accidents. In this paper, we propose a novel Sybil attack detection
%mechanism, Proofprint, using the trajectories of vehicles as its identities. Each RSU is responsible for issuing a time-stamped tag as a proof of its appearance.  We design a \textit{proof of location} authorized message for two objectives: First, not only one RSU is able to create a trajectory for the vehicle, rather, vehicles should travel a long a certain number of RSUs; Second, to establish trust relationship between trajectories, two signed messages within the same period of time can be recognized. In addition, to limit the ability of vehicles to have  multiple trajectories, upon receiving the proof of location message from RSU, the vehicle should use it as a seed to run the \textit{proof of work algorithm} to provide a proof for the next RSU that she used its computational power in calculating such a proof. Our extensive simulations and real-worldexperiments demonstrate that Proofprint achieve high detection rate with low possibility of false negatives. Additionally, the scheme shows acceptable communication and computation.

Vehicular Ad Hoc Networks (VANETs) has the potential to enable the next-generation Intelligent Transportation Systems (ITS). In ITS, data contributed from vehicles can build a spatiotemporal view of traffic statistics, which can consequently improve road safety and reduce slow traffic and jams. To preserve vehicles' privacy, vehicles should use multiple pseudonyms instead of only one identity. However, vehicles may exploit this abundance of pseudonyms and launch Sybil attacks by pretending to be multiple vehicles. Then, these Sybil (or fake) vehicles report false data, e.g., to create fake congestion or pollute traffic management data. In this paper, we propose a Sybil attack detection scheme using proofs of work and location. The idea is that each road side unit (RSU) issues a signed time-stamped tag as a proof for the vehicle's anonymous location. Proofs sent from multiple consecutive RSUs is used to create vehicle trajectory which is used as vehicle anonymous identity. Also, one RSU is not able to issue trajectories for vehicles, rather the contributions of several RSUs are needed. By this way, attackers need to compromise an infeasible number of RSUs to create fake trajectories. Moreover, upon receiving the proof of location from an RSU, the vehicle should solve a computational puzzle by running proof of work (PoW) algorithm. So, it should provide a valid solution (proof of work) to the next RSU before it can obtain a proof of location. Using the PoW can prevent the vehicles from creating multiple trajectories in case of low-dense RSUs. Then, during any reported event, e.g., road congestion, the event manager uses a matching technique to identify the trajectories sent from Sybil vehicles. The scheme depends on the fact that the Sybil trajectories are bounded physically to one vehicle; therefore, their trajectories should overlap. Extensive experiments and simulations demonstrate that our scheme achieves high detection rate to Sybil attacks with low false negative and acceptable communication and computation overhead.
	\end{abstract}

\section{Introduction} 
\label{sec:introduction}
% VANETS
	 
Over the last two decades, Vehicular Ad Hoc Networks (VANETs) have been emerging as a cornerstone to the next generation Intelligent Transportation Systems (ITSs), contributing to safer and more efficient roads. In VANETs, moving vehicles are enabled to communicate with each other via intervehicle communications as well as with road-side units (RSUs) in vicinity via RSU-to-vehicle communications. As a result, a wide spectrum of applications have been emerged as promising solutions~\cite{wu2014urbanmobilitysense} to enable new forms of ubiquitous traffic management applications that are not possible with our current traditional transportation system. The core idea of these applications is to enable vehicles to contribute with data and feedback to an event manager which can build a spatiotemporal view of the traffic state and also to extract important jam statistics~\cite{hu2015smartroad}. These applications have the potential to contribute to safer and more efficient roads by enabling a wide range of applications such as pre-crash sensing and warning, traffic flow control, local hazard notification, and enhanced route guidance and navigation~\cite{rabieh2015cross}.

However, the aforementioned applications depend on information sent from participating vehicles. Therefore, it is required to preserve drivers privacy especially location privacy while still verifying their identities in an anonymous manner~\cite{chang2012footprint, machardy2018v2x}. A naive solution is to allow each vehicle to have a list of pseudonyms to be authenticated anonymously. However, a malicious vehicle may abuse this privacy protection to launch \textit{Sybil attack}~\cite{qu2015security}. In Sybil attacks, a malicious vehicle uses its pseudonyms to pretend as multiple fake (or Sybil) nodes~\cite{reddy2017sybil}. The consequences of a Sybil attack in VANETs can be disastrous. For example, a malicious vehicle can launch the attack to create an illusion of traffic congestion. Consequently, other vehicles will choose an alternative route and evacuate the road for the malicious vehicle. Another potential consequence of a Sybil attack is in safety-related applications such as collision avoidance and hazard warnings where a Sybil attack can lead to biased results that may result in car accidents~\cite{rabieh2015cross}. Hence, it is of great importance to detect Sybil attacks in VANETs.

Existing works of detecting Sybil attacks can be categorized into three categories, namely, identity registration, position verification and trajectory-based approaches. The ultimate goal of these detection mechanisms is to ensure each physical node is bounded with a valid unique identity. Firstly, identity registration approaches~\cite{zhou2011p2dap, reddy2017sybil, el2011privacy} require a dedicated vehicular public key infrastructure to certify individual vehicles with multiple pseudonyms to ensure each physical node is bounded with a valid unique identity. However, identity registration alone cannot prevent Sybil attacks, because a malicious node may get multiple identities by non-technical means such as stealing or even collusion between vehicles~\cite{yao2018multi}. Secondly, position verification approaches depend on the fact that individual vehicle can present at only one location at a time. In~\cite{bouassida2009sybil},~\cite{rabieh2015cross}, localization techniques such as Global Positioning System (GPS) are used to provide location information of vehicles to detect Sybil nodes. However, these schemes fail due to the highly mobile context of vehicular networks~\cite{syed2004fuzzy}. Thirdly, trajectory-based approaches is based on the fact that individual vehicles move independently, and therefore they should travel along different routes. In~\cite{chang2012footprint}, the vehicle obtains its trajectory by combining a consecutive tags from RSUs which it encounters. However, the scheme suffer RSU compromise attack in which if one RSU is compromised, a malicious vehicle can obtain infinite number of valid trajectories. Moreover, in case of rural areas (RSUs are not dense), attackers can create valid trajectories that look for different vehicles. %are free to start a trajectory any time by simply generating a temporary public/private key pairs. %Thus, a malicious vehicle may abuse this freedom by asking for multiple trajectories to create many forged trajectories to launch a Sybil attack.

In this paper, we propose a novel Sybil attack detection scheme using proofs of work and location. The main idea is that when a vehicle encounters an RSU, the RSU should issue authorized \textit{time-stamped tag} which is a concatenation of time of appearance and anonymous location tag of that RSU. As the vehicle keeps moving, it creates its \textit{trajectory} by combining a set of consecutive authorized time-stamped tags that are chronologically chained to each other. That trajectory is used as an anonymous identity of the vehicle. Since RSUs have the main responsibility to issue proof of location to vehicles, the scheme should resist against \textit{RSU compromise} attack so we design the trajectory so that not only one RSU is capable of creating trajectories for the vehicles. To achieve this, threshold signature is adopted so that each RSU is only able to generate a partial signature on a set of time-stamped tags. Once a vehicle travels along a certain threshold number of RSUs, a standard signature representing a proof of location can be generated. Upon receiving an authorized message from an RSU, the vehicle should use it as a seed to solve a puzzle using a \textit{proof-of-work} algorithm, similar to the one used in \textit{Bitcoin}~\cite{nakamoto2008bitcoin}. The core idea of PoW is to provide a proof to RSUs so they can ensure that the vehicle solved the puzzle correctly. Comparing to Footprint~\cite{chang2012footprint}, using PoW limits the ability of a malicious vehicles to create multiple trajectories. To detect Sybil trajectories, upon receiving an event from other vehicles, the \textit{event manager} first applies a set of heuristics to construct a connected graph of Sybil nodes, then it uses the maximum clique algorithm~\cite{tomita2010simple} to detect all Sybil nodes in that graph.

Our main contributions and the challenges the paper aims to address can be summarized as follows:
    \begin{itemize}

        \item  We used threshold signatures to resist RSU compromise attacks. The attacker needs to compromise an infeasible number of RSUs to be able to create fake trajectories.

    \item We used the PoW algorithm to limit the ability of a malicious vehicle to create multiple forged trajectories, and more importantly, to reduce the detection time for detecting Sybil trajectories which is a critical concern in traffic management applications.
        
    \item We carefully analyzed the probabilistic nature of PoW based scheme by examining the affecting parameters (e.g travel time between two consecutive RSUs) experimentally, and then we developed a mathematical model that can be used for adjusting these parameters so that the ability of a malicious vehicle to create forged trajectories is reduced significantly. 
        
        %\item In the proposed scheme, each RSU should be preloaded with a Look-up target table that contains targets corresponding to traverse time of vehicle between two adjacent RSUs,  we provide detailed experimental method followed by a mathematical model for the selection of the target value used by the proof of work algorithm.

        \item By experiments, we prove that using the proof of work algorithm reduces the ability of a malicious vehicle to maintain actual multiple trajectories simultaneously. Further simulations, analysis, and practical experiments are conducted to evaluate the proposed scheme and compare it with the Footprint~\cite{chang2012footprint}, the results indicate that the proposed scheme can successfully detect and defend against Sybil attacks in VANETs and more efficiently compared to the Footprint.

        %\item Simulations, analysis, and practical experiments are conducted to evaluate the proposed scheme. The results indicate that the proposed scheme can successfully detect and defend against Sybil attacks in VANETs efficiently compared to the Footprint. 

 \end{itemize}
    
        The rest of the paper is organized as follows. We describe the network and threat models in VANETs, followed by the design goal of our Sybil detection scheme in Section~\ref{model and design goal}. In Section~\ref{Preliminaries}, we discuss preliminaries used by this research work. Then, our proposed scheme is presented in Section~\ref{sec:ProposedScheme}. In Section~\ref{Selection of PoW targets}, we show the selection of PoW parameters values experimentally, and also we provide a mathematical proof of the experimental results. Detailed security and performance evaluations are provided in Section~\ref{performance}. We present the computation complexity analysis of our scheme in Section~\ref{Computaion overhead}. Section~\ref{related} discusses the previous research work in Sybil detection in VANETs. Finally, we give concluding remarks in Section~\ref{conclusion}.

	\section{Models And Design goals}
	\label{model and design goal}
    In this section, we present the considered network model followed by the adversary and threat models, and then, we introduce the design goals of our scheme.

   \subsection{Network Model}
\begin{figure}[tp]
			\centering
			
			\includegraphics[width=1\linewidth]{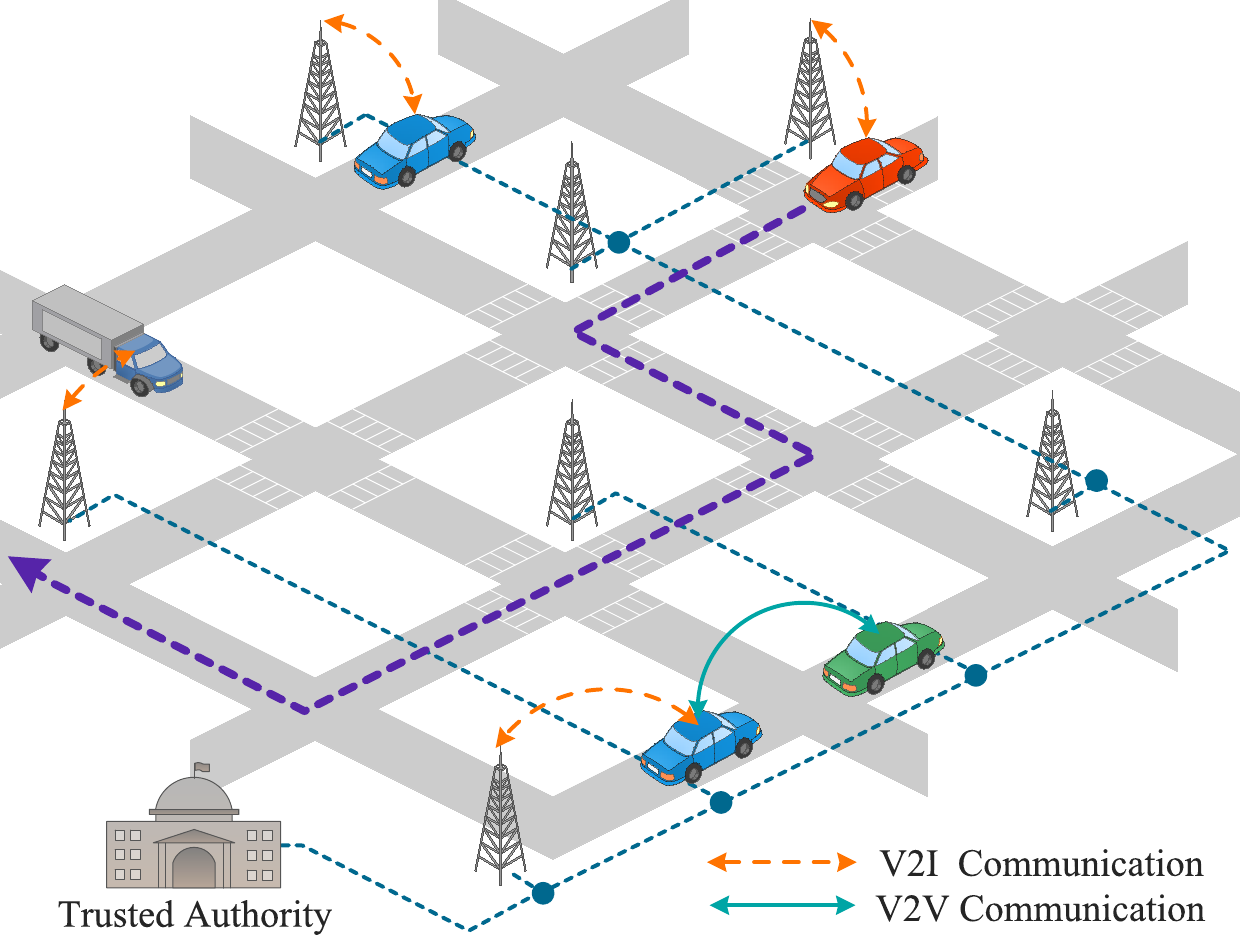}
			\vspace{0mm}
			
			\caption{The considered network model where the purple dash line indicates the travel path of a vehicle in which a number of RSUs are encountered.}
			\label{model1}
		\end{figure}

    As depicted in Fig.~\ref{model1}, the considered network model has the following entities.
   
    \begin{itemize}
        \item \textbf{Roadside units (RSUs):} RSUs act as a typical wireless access points that can communicate with vehicles within its vicinity. They can communicate with each other via a dedicated network or the Internet~\cite{chang2012footprint}. %Typically, RSUs are deployed at road intersections or any area of interest.
        
        \item \textbf{Vehicles:} Vehicles are equipped with On-board units (OBUs) that have two main parts: a short-range wireless module (e.g., DSRC IEEE 802.11p~\cite{alsabaan2013vehicular}) and a GPS receiver. Vehicles communicate with each other and with RSUs that are deployed along the road.

        \item \textbf{Offline Trusted Authority (TA):} TA is responsible for vehicles registration, issuing digitally certified pseudonyms to vehicles, deployment of RSUs, and ensuring the security of VANETs. In practice, the TA can be the Department of motor vehicles (DMV). 
        
        \item \textbf{Event Manger (EM):} EM is responsible for managing the traffic management applications and also detecting Sybil trajectories upon receiving a reported event from vehicles.

        %In our scheme, the TA is not involved in issuing certificates for vehicles since vehicles do not have to use explicit identities.     

%\item   \textbf{Event Manger (EM):} EM is a dedicated entity which is responsible for detecting Sybil trajectories upon receiving a reported message from vehicles. 
		
	\end{itemize}

	    %In this paper, we make the following assumptions \textit{Vehicles move independently,} meaning that they do not have the exact same routes for all the time. And \textit{the RSUs are synchronized,} synchronization of RSUs is easy to accomplish since all RSUs are interconnected via the RSU backbone network.

	\subsection{Adversary and Threat Model}
	
 The TA is fully trusted because it is operated by the government which is interested in the security of the VANET. RSUs are \textit{honest-but-curious} in running the scheme correctly but they are interested to collect the location information of the drivers. Also, attackers can compromise a number of RSUs. The vehicles are not trusted and may launch Sybil attacks by pretending as multiple fake identities. Attackers can collude with each other and share their pool of pseudonyms. The event manager is \textit{honest-but-curious} in running the Sybil attack detection scheme, but it should not be able to reveal the actual identities of drivers.

	\subsection{Design Goals}
	
	Our Sybil detecting scheme should achieve the following objectives: 
	
	\begin{enumerate}
		
			\item \textit{Resisting RSU compromise.} Since RSUs are the responsible for generating trajectories for vehicles, the scheme should be resistant to  RSU compromise attack.
			
			%\item \textit{Efficient trajectory Creation:} The design of our scheme should allow honest vehicles to have one valid trajectory. Meanwhile, mitigate other malicious vehicles' ability to obtain multiple trajectories.  
			
			\item \textit{Short Sybil attack detection time.}  The time to  detect Sybil attacks should be short. This objective is important in time-sensitive applications such as traffic management.

			%mOnce an adversary launches a Sybil attack, the detection scheme should react before the attack ends i.e., the detection time of Sybil nodes should be very small. Otherwise, the adversary could already accomplish its objective. %This is critical especially in safety related applications in VANETs that require the event manager to be able to detect Sybil nodes instantaneously. 
			
		\item \textit{Location privacy preserving.} The location privacy of vehicles' drivers should be preserved while vehicles need to authenticate themselves anonymously. %manner. The detection scheme should prevent the leakage of information concerning the location of vehicles.
		
		\item \textit{Unlinkability.} The message sent by a vehicle at different locations or times should not be linkable. Also the messages of the same vehicle should not be linkable over the time. This Unlinkability objective is important because $(i)$ by knowing the locations visited by anonymous drivers, the attacker can identify the driver; $(ii)$ by linking messages, too much location information can be exposed if the attackers manages to know the sender of the message.

		%The trajectories of the same vehicle should not linkable over the time. Since linking trajectories enables tracking vehicles and collecting too much information on vehicles  so that in case the attackers were able to link a trajectory to a specific sender, alot of information on the sender can be collected.   

	\end{enumerate}

		\section{Preliminaries}
		\label{Preliminaries}
		
		In this section, we present the necessary background on secret sharing and threshold signatures that we will use in our scheme. The notation  used in the paper are listed in Table.~\ref{not}
				
		\begin{table}[!t]
			\centering
			\caption{System Notations.}
			\begin{tabular}{l|l}
				\hline \hline
				Symbol                                       & Description \\ \hline
				$t$ & The threshold of $(t, n)-$threshold signature protocol.\\ \hline
				$SK_{R_i}$ &  The partial private key assigned to the RSU $R_i$  \\ \hline
				$PK_{v_{i}}/ SK_{v_{i}}$                     & Public/ Private key pair for vehicle $V_i$. \\ \hline

				$\sigma_{R_{i}}$ & A partial signature from RSU $R_i$.\\ \hline
				
				 $C_{TA}(PK^{j}_{v_i})$ & A certificate from the TA on $PK^{j}_{v_i}$\\ \hline
				 $T$ & A target from solving the PoW puzzle. \\ \hline
				 $\mathcal{T}_i$ & The trajectory of a vehicle.  \\ \hline

				 $l$ & Number of RSUs that a vehicle encounters.\\ \hline

			\end{tabular}
			\label{not}
		\end{table}
		\subsection{Secret Sharing}
		In a secret sharing schemes, a trusted party distribute a secret key $d$ among a group $\mathcal{P}=\{P_1,...,P_n\}$ of $n$ participants, each of which owns a share of the secret in such a way that at least any $t$ participants can use their secret shares to reconstruct the secret.
		
		Shamir~\cite{shamir1979share} have proposed a $(t, n)$-threshold secret sharing scheme in which at least $t$ participants are capable of recovering the secret. Indeed, let $\mathbb{Z}_{p}$ be a finite field with $p$ > $n$ and $d \in \mathbb{Z}_{p}$ is the secret key to be shared. The TA chooses a polynomial $q(x)$ of degree at most $t-1$ which can be written as:

		$$q(x)= d +\sum_{j=1}^{t-1} {a_{j}x^{j}}$$
		
		where $a_{j} \in \mathbb{Z}_{q}$ is randomly chosen. Each participant ($P_{i}$) is assigned a secret share $d_{i}=q(\alpha_{i})$. The set of shares yields a $(t, n)$ threshold access structure where at least a set $A\in \mathcal{P}$ can retrieve the secret key $d$ using  Lagrange interpolation technique as follows:

		$$d= \sum_{P_{i} \in A}d_{i} \delta^{A}_i = \sum_{P_{i} \in A}d_{i}(\prod_{P_{j}\in (A\textbackslash P_{i})}\frac{-\alpha_{j}}{\alpha_{i}-\alpha_{j}})$$
		
		Values $\delta^{A}_i$ are known as Lagrange coefficients. It can be proven that less than $t$ participants cannot get the secret $d$.

		\subsection{Threshold signatures}
		A regular signature generation algorithm takes as inputs message, $m$, and a sender's private key $SK$ and output a digital signature $\sigma_{SK}(m)$. The receiver can verify the signature using the sender's public key $PK$. A (t, n)-threshold signature is used to share the signing operation between a subset $t$ of $n$ participants rather than giving the power of signing to only one participant. The idea is that a secret key, $SK$, is divided into shares and each share ($SK_i$) is assigned to one of the group participants. To sign a message, a member can use his secret share of the secret key to generate a partial signature called signature share $\sigma_i$. Then, a subset of at least $t$ participants can compute a valid signature $\sigma_{SK}(m)$ on $m$ by combining their signature shares. This signature can be verified by anybody using a unique public key.
		
We adopt an efficient threshold signature scheme proposed by Alexandra~\cite{boldyreva2003threshold} that is based on Gap Diffie-Hellman (GDH) groups~\cite{boneh2001short} for forth reasons. First, it has proven to be secure. Second, anonymity is provided by unlinking the $SK_{i}$ with the identity of the signer. Third, generating a signature does not require any interaction or any zero-knowledge proofs with versifiers. Forth, the signature shares are short, and signature reconstruction requires only the multiplication of signature shares. Moreover, it imposes low computational overhead since the signing process only requires hash computations and modular exponentiation and the verification process requires two pairing operations. 
		
A brief description to the threshold signature scheme is as follows. Let $\mathbb{G}$ be a GDH group of prime order $p$ and $g$ be a generator of $\mathbb{G}$ and $\mathcal{H}:\{0, 1\}^{*}\rightarrow{\mathbb{G}}$ is a public one-way and collision-resistant hash function. Each participant $P_{i}$ should have a secret share $SK_{i}$ using methods described in~\cite{gennaro}. A participant $P_i$ uses its secret share to compute the signature share on a message ($m$) as: $$\sigma_{SK_{i}}(m) = \mathcal{H}(m)^{SK_{i}} $$

		After a set $A$ of at least $t$ participants compute their signature shares for message $m$, a standard signature for the message can be calculated as:
		\begin{equation}
		\label{threshold}
		\sigma_{SK}({m}) = \prod_{i\in \mathcal{A}}\sigma_{SK_{i}}(m)^{\delta^{A}_i} = \mathcal{H}(m)^{\sum _{i\in{\mathcal{A}}} \delta^{A}_i SK_i} = \mathcal{H}(m)^{SK}
		\end{equation}
		
		Where $\delta^{A}_i$ are Lagrange coefficients.

		\section{Proposed scheme}
		\label{sec:ProposedScheme}
		In this section, we present our Sybil detection scheme. We first start with system initialization, then we show how vehicles obtains proof-of-location messages. Then, we describe the role of the PoW followed by an illustrative example of trajectory creation in our scheme. Finally, we describe the how the event manager can detect Sybil trajectories.

	\subsection{Overview}

In our scheme, vehicles should request proof of location form each RSU it encounters as a proof of its presence there. The issued proof of location message should be \textit{temporarily linkable} to preserve the drivers' locations privacy.

In our scheme, we used threshold signatures to prevent such kind of attack. We introduce \textit{Proof-of-Work (PoW)} algorithm as \textit{first layer of defence} to Sybil attacks since it reduces the possibility of a malicious vehicle to obtain multiple trajectories successfully. The vehicle should solve a puzzle while moving to the next RSU. Then, an RSU can issue a proof of location to the vehicle but after verifying the puzzle solution. However, some malicious vehicles can create multiple trajectories to launch a Sybil attack. Therefore, we apply a \textit{second layer of defence} using some heuristics that define forged trajectories (created by malicious vehicles). So, once a vehicle reports a particular event such as road congestion or accident, it should also submit its trajectory (A set of consecutive proof of locations issued for a vehicle by RSUs that are timely chained together) for identification. Then, a event manager conducts Sybil attack detection by first check the similarity relationship among each pair of trajectories. Then, Sybil trajectories from the same attacker are overlapped within the same "group". Finally, each group will be considered as one single physical vehicle. By this way, Sybil nodes can be eliminated. In the following sections, we describe our scheme in details.

\subsection{System Initialization}
During this stage, RSUs are divided into groups in such a way that each group cover a certain area or a road segment. Then, the TA sends to the RSUs within each group the credentials they need for a $(t,n)$ threshold signature, where $n$ is the number of RSUs deployed in the group. In order to do this, a polynomial of degree $t-1$ is calculated at different points $\alpha_{i}$, for $i=1$ to $n$. Then, the TA generates a public key $PK$ and a list of $n$ secret key shares $SK_{i} \, \forall \, i=\{1,\dots,n$\} of the secret key ($SK$). After that, TA sends to each RSU via a secure channel a secret key share $(SK_i)$ and the public key $(PK)$ as well as a list of public keys corresponding to the secret key shares of the neighbouring RSUs. To ensure RSUs are issuing trajectories for legitimate vehicles, each vehicle generates a set of public/private key pairs and obtain certificates for each public key from the TA so that vehicles can anonymously authenticate themselves to the RSUs. Getting these certificates from the TA can be done during vehicle registration from, e.g., the Department of motor vehicles (DMV). 
        
%Also, each RSU should in advance preloaded with a \textit{look-up target table} necessary for the operation of PoW algorithm. Details of how to obtain the look-up table will be discussed later in Section.~\ref{Selection of PoW targets}. For a vehicle to join in the system, it only needs to get the public key list of RSUs that it encounters and the public key $(PK)$.

		%One possible design to generate authorized messages that define trajectory for a vehicle is that each RSU periodically broadcasts authorized time stamped tags to the vehicles in its coverage. However, any vehicle may claim its appearance at a particular RSU by getting such tags via eavesdropping on the wireless channels. Therefore, time stamped tags should be issued for vehicles individually.
		
		\begin{figure*}[!t]
				\centering
				
				\includegraphics[width=1\linewidth, frame]{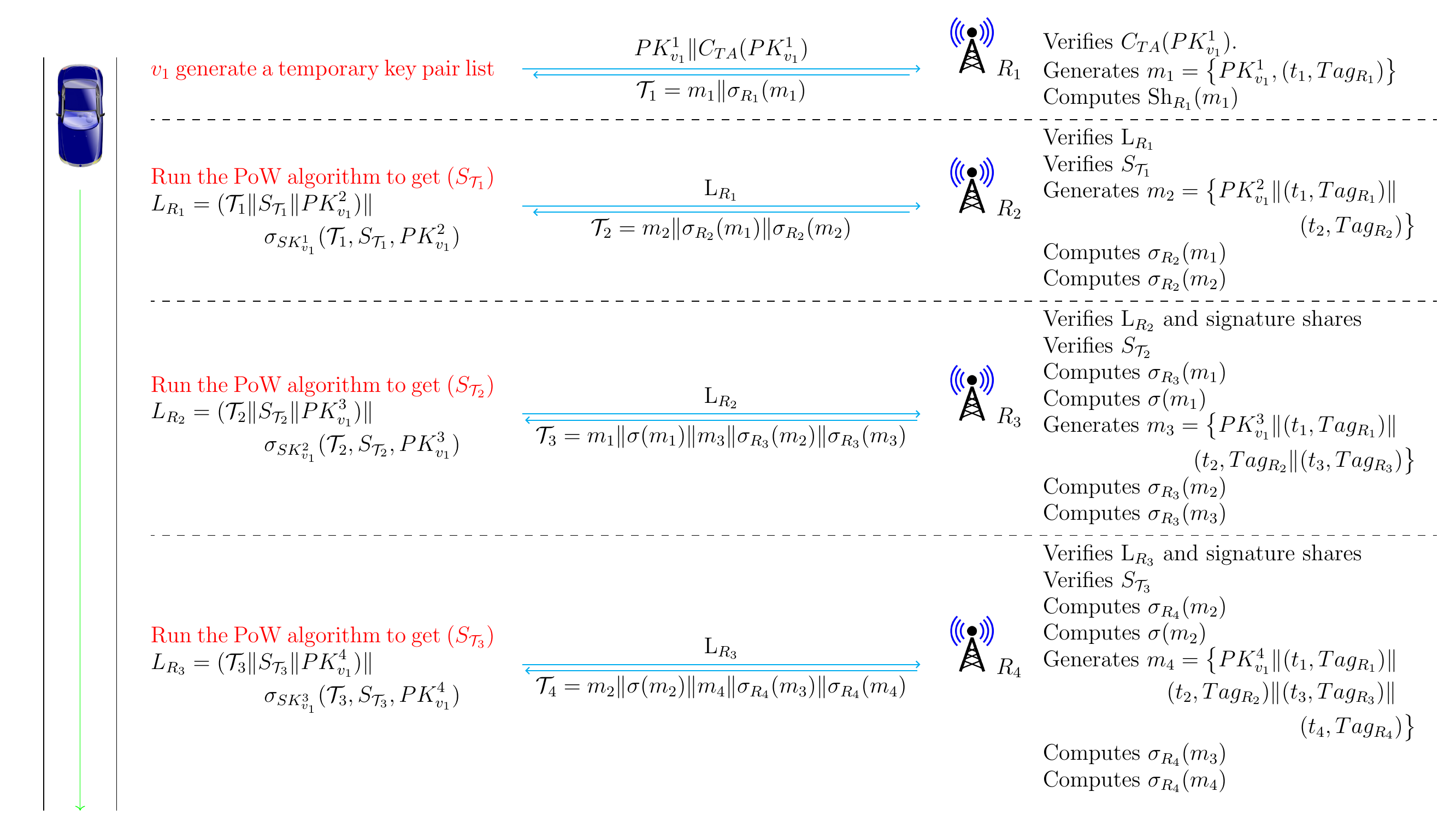}
				\vspace{0mm}
				
				\caption{Exchanged messages in our scheme with threshold of $t=3$.}
				\label{proposed}
			\end{figure*}

		\subsection{Exchanged Messages}

In this subsection, we illustrate by an example of how a vehicle can construct its trajectory through Fig.~\ref{proposed}. Assume a threshold signature with $t=3$ and a vehicle $v_{1}$ is traveling along the following RSUs [$R_{1}, R_{2},R_{3},R_{4}$]. To ensure anonymity, $v_{1}$ should in advance generate a list of temporary public/private keys. Then, $v_1$ starts creating its trajectory as follows:
		
		\begin{enumerate}
			
\item When $v_1$ first encounters $R_{1}$, it can start its trajectory by simply sending a public key $PK^{1}_{v_{1}}$  from the temporary key pair $[PK^{1}_{v_{1}}, SK^{1}_{v_{1}}]$ with the certificate of the TA $C_{TA}(PK^{1}_{v_1})$.

			 \item $R_{1}$ first authenticates the vehicle if it is legitimate to request a proof of location by checking the vehicle's certificate. Then, it generates a proof of location message $m_1 = \{PK^{1}_{v_{1}}, (t_1, Tag_{R_{1}})\}$ where $t_{1}$ is the current time stamp and $\sigma_{R_{1}}(m_{1})$ is signature share of $R_1$ on $m_{1}$. Then, $R_{1}$ sends $\mathcal{T}_{1}= m_{1} || \sigma_{R_{1}}$ back to the vehicle.

			\item Upon receiving the message ($\mathcal{T}_{1}$) and to limit a malicious vehicle ability to generate multiple trajectories, the vehicle should use it as a seed to run the \textit{proof of work} (PoW) algorithm. The idea of using Proof-of-work concept, was proposed in~\cite{back2002hashcash} to defend against denial-of-service attacks and email spams, and recently has become more popular for its use to secure bitcoin~\cite{nakamoto2008bitcoin}. In PoW, a \textit{prover} (i.e., a vehicle) before request a proof of location from an RSU, he/she should solve a certain computational challenging puzzle to prove that a certain amount of time has been taken to solve this puzzle. The \textit{verifier} (i.e., an RSU) on the other side can verify that the proof is valid with a \textit{negligible computational cost}. We adopt the \textit{Hashcash} PoW technique used in \textit{Bitcoin}~\cite{nakamoto2008bitcoin} as follows:
	
	\begin{equation}
		\label{PoW}
		H(n||H(\mathcal{T}_{R_1})) < T
	\end{equation}

where $H$ is a secure hash function e.g. SHA-256, $\mathcal{T}_{R_1}$ is the authorized proof of location issued from an RSU $R_1$, $T$ is an expected target value and $n$ is a nonce value. The puzzle problem is to find the appropriate nonce value $n$ to make the hash value less than a certain target, $T$. Brute force is the only known way to solve the puzzle problem. The computational difficulty of problem solving depends on the value of $T$. The smaller the value $T$, the more  difficult it is to generate the proper nonce value and more time is required to generate such nonce.			
%The RSU should issue a unique seed value for each vehicle to be used to solve the PoW puzzle. That is required to ensure that the vehicles can not exchange the solution of the puzzle among each other.Since $\mathcal{T}_{R_j}$ is unique for each vehicle, an authorized message issued by an RSU can be used to derive a unique seed value for PoW algorithm described in Eq.~\ref{PoW}.
Note that the longer it takes a vehicle to travel between two RSUs, the more time it has to solve the PoW puzzle and hence the lower the expected target value should become. Upon arriving at the next RSU, the vehicle should submit the lowest $T$ value resulted from running the PoW algorithm to the RSU. To illustrate, after a vehicle $v_1$ receives the authorized message $\mathcal{T}_{1}$ from $R_{1}$, it should use $\mathcal{T}_{1}$ as a seed value to run the PoW algorithm. Once $v_1$ encounters the next RSU (i.e., $R_{2}$), it should first generate a new temporary public/private key pair $PK^{2}_{v_{1}}, SK^{2}_{v_{1}}$ and then sends the solution of solving the PoW puzzle denoted by $S_{\mathcal{T}_{1}}$ included in following message to $R_2$:
$L_{R_{1}}=(\mathcal{T}_{1}||S_{\mathcal{T}_{1}}||
			PK^{2}_{v_{1}} ) || \sigma_{SK^{1}_{v_{1}}}(\mathcal{T}_{1}, S_{\mathcal{T}_{1}}, PK^{2}_{v_{1}})$

\item Once $R_{2}$ receives $L_{R_{1}}$ from $v_1$, $R_{2}$ verifies the message $L_{R_2}$ in two steps:
\begin{enumerate}
\item \textit{Ownership verification}: $R_2$ first take  $PK^{1}_{v_{1}}$ from $\mathcal{T}_{1}$, and check whether $V_{PK^{1}_{v_1}}(\sigma_{{SK^{1}_{v_{1}}}}(\mathcal{T}_{R_1},S_{\mathcal{T}_{R_1}}, PK^{2}_{v_{1}}))= \mathcal{T}_{R_1} ||S_{\mathcal{T}_{R_1}} || PK^{2}_{v_{1}}$. This is mandatory so that authorized messages cannot be misused by other vehicles since only the vehicle $v_1$ knows $SK^{1}_{v_{1}}$ that corresponds to its pairwise $PK^{1}_{v_{1}}$. Furthermore, $R_{2}$ further examines whether the signature share contained in $\sigma_{R_{1}}$ is signed by one of its neighboring RSUs.
			
\item \textit{PoW puzzle verification:} Once $L_{R_{1}}$ succeeds the ownership verification step, $R_{2}$ further check the validity of PoW puzzle solution ($S_{\mathcal{T}_{R_1}}$) by $(i)$ computing $H(S_{\mathcal{T}_{R_1}}||H(\mathcal{T}_{R_1}))$ $(ii)$ Determining the travel time of $v_1$:= $t_2-t_1$ where the time stamp $(t_1)$ is taken from $\mathcal{T}_{R_1}$ and $t_2$ is the current time of receiving $L_{R_{1}}$. Thereafter, it uses a look up target table to check whether the target corresponding to the time taken by $v_1$ to travel between $R_{1}$ and $R_{2}$ is actually below a certain target value. If the verification of $S_{\mathcal{T}_{R_1}}$ does not meet the required difficulty, an $R_{2}$ refuses contributing by its signature share to the corresponding trajectory, forcing the vehicle to start over with a new trajectory.
\end{enumerate} Note that if a vehicle does not pass either the above steps, such a vehicle will be considered as a malicious vehicle, and the RSU will terminate any further communications with it. It is worth mentioning that even if a malicious vehicle passes the \textit{ownership verification} step and due to the limitation of a vehicle's computational resources, it would be difficult for it to generate multiple trajectories simultaneously because it would have to solve a separate puzzle for each authorized message it obtains from an RSU.
	
Then, $R_2$ generates $m_{2}=PK^{2}_{v_{1}} || (t_{1}, Tag_{R_{1}}) || (t_{2}, Tag_{R_{2}})$ and sends $\mathcal{T}_{2}= \{m_{2} || \sigma_{R_{2}}(m_{1}) || \sigma_{R_{2}}(m_{2} )\} $ back to $v_{1}$. Note that $\sigma_{R_{2}}(m_{1})$ and $\sigma_{R_{2}}(m_{2})$ are signature shares of $R_{2}$ on $m_{1}$ and $m_{2}$ respectively.

			%$R_{2}$ uses $PK^{1}_{v_{1}}$ contained in $\mathcal{T}_{1}$ to verify $L_{R_{1}}$. Then it verifies the correctness of PoW puzzle solution ($S_{\mathcal{T}_{1}}$) by first calculating the travel time of $v_1$ which is the time deference between receiving $L_{R_{1}}$ and $t_1$ time of appearance at $R_1$. Then, it checks whether the travel time corresponds to the received target value in its look up target table. Also, it checks for whether the signature share ($\sigma_{R_{1}}$) is from a neighboring RSU or not. Then, it generates $m_{2}=PK^{2}_{v_{1}} || (t_{1}, Tag_{R_{1}}) || (t_{2}, Tag_{R_{2}})$ and sends $\mathcal{T}_{2}= \{m_{2} || \sigma_{R_{2}}(m_{1}) || \sigma_{R_{2}}(m_{2} )\} $ back to $v_{1}$. Note that $\sigma_{R_{2}}(m_{1})$ and $\sigma_{R_{2}}(m_{2})$ are signature share of $R_{2}$ on $m_{1}$ and $m_{2}$ respectively.  

		\item As this vehicle moves on and encounters
			$R_3$, it sends $L_{R_{2}}$. The authorized message generation process at $R_{3}$ is similar to $R_2$ but $R_{3}$ further use the threshold signature scheme to compute a standard signature $\sigma({m_{{1}}})$ for $m_{1}$. Then, $R_{3}$ sends $\mathcal{T}_{3}= {m}_{1} || \sigma({m_{{1}}})|| m_{3} || \sigma_{R_{3}}(m_{2}) || \sigma_{R_{3}}(m_{3} )$ back to the vehicle, where $m_{3}=\{PK^{3}_{v_{1}} || (t_{1}, Tag_{R_{1}}) || (t_{2}, Tag_{R_{2}})\\ ||(t_{3}, Tag_{R_{3}})\}$.
			
	\end{enumerate}
	
	 Note that $m_{1} || \sigma({m_{{1}}})$ represents the proof of location for the appearance of the vehicle at $R_{1}$ at time $t_{1}$. Hereafter, the vehicle can start creating its trajectory.

	\subsection{Creating Trajectories and Reporting Events}	
	The vehicle can start creating its own trajectory after she encounters $t$ RSUs. As in Fig.~\ref{proposed}, upon $v_1$ reaches at $R_{4}$, after it checks whether the received message $L_{R_{3}}$ passes the ownership and PoW verification steps, it can generate a standard signature $\sigma({m_{2}})$ over $m_{2}$. Here, $m_{2} || \sigma({m_2})$ represents a proof of location of $v_1$ at both $R_1$ and $R_2$. The process is repeated such that $R_{4}$ sends $\mathcal{T}_{4}= {M}_{2} || \sigma({m_{{2}}}) || m_{4}||\sigma_{R_{4}}(m_{3})|| \sigma_{R_{4}}(m_{4} )$ back to the vehicle, where $m_{4}=\{PK^{4}_{v_{1}} || (t_{1}, Tag_{R_{1}}) || (t_{2}, Tag_{R_{2}} ) ||\\ (t_{3}, Tag_{R_{3}} || (t_{4}, Tag_{R_{4}})\}$.

As the vehicle moves on, a set of consecutive time stamped authorized location tags issued for a vehicle are tightly chained together to form a trajectory of a vehicle. That trajectory is used as the vehicle unique anonymous trajectory.

\subsection{Detecting Sybil Attacks}

\label{sec:DetectingSybilAttacks}
In this subsection, we explain how the event manager can detect and eliminate Sybil attacks. When a vehicle reports an event (e.g. an accident or a traffic jam) to the  event manager, the vehicle should send its trajectory with the event message. However, since the PoW algorithm limits (but not eliminate) the chance for a malicious vehicle to obtain multiple trajectories due to its probabilistic nature as will be discussed in details in Section.~\ref{Selection of PoW targets}, we use a set of heuristics and graph based representations as \textit{a second layer of defence} to recognize and remove trajectories sent from Sybil nodes. Note that the use of PoW is the \textit{first line of defense}. Basically, and similar to~\cite{chang2012footprint}, there are two main heuristics that can be used to decide if two trajectories are distinct (created by real vehicles) or forged (created by a malicious vehicle): 

\begin{enumerate}
	\item \textit{Check window size/Traverse time limit} is the time required for a vehicle to travel between two consecutive RSUs. This heuristic is used since a single vehicle should not be able to traverse two consecutive RSUs in a time shorter than a certain limit that is traverse time limit or check window size.  
	
	\item  \textit{Trajectory length limit} is defined as the maximum number of RSUs traversed by a single vehicle within a time period. This heuristic is used since in practice vehicle will not be able to exceed a certain limit of RSUs traversed in a particular time. 
	\end{enumerate}

\begin{figure}[!t]
			\centering
			
			\includegraphics[width=1\linewidth]{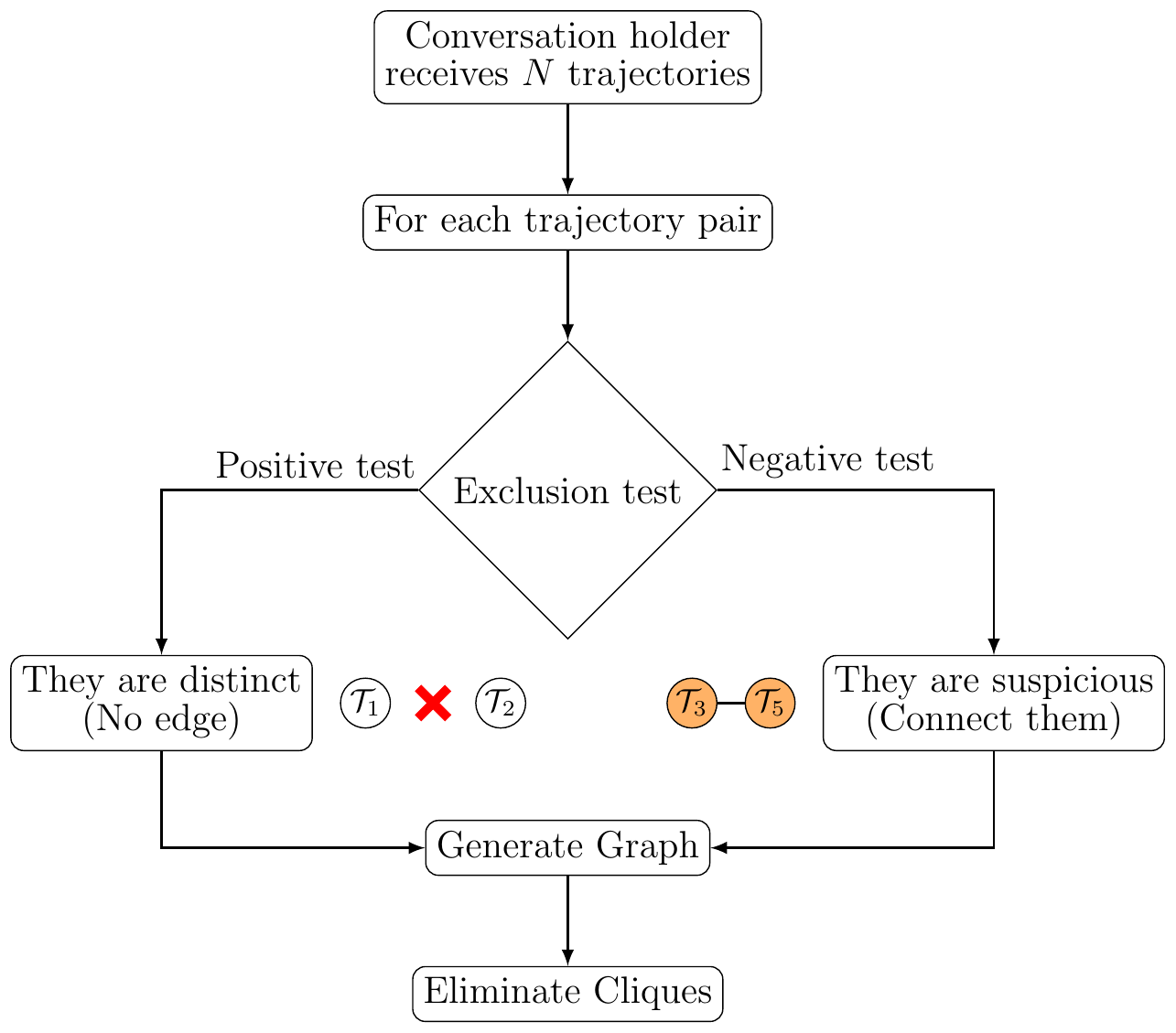}
			\vspace{0mm}
			\caption{Flow chart of detecting Sybil nodes.}
			\label{uml}
		\end{figure}

Based on these two heuristics, upon receiving a number of $N$ trajectories by event manger, he/she can conduct Sybil attack detection as shown in flow chart illustrated in Fig.~\ref{uml} as follows:
	 	 	
\textbf{Phase 1: Conducting exclusion test.} In this phase, the event manger conduct an \textit{exclusion test} for each pair of received trajectories. Mainly, there are two cases where a pair of trajectories can pass the test (\textit positive test). First, if there are two distinct RSUs tags appearing in the same check window. Second, if the number of RSUs obtained by combining all distinct RSUs in two trajectories is greater than the trajectory length limit. In all other cases, the pair of trajectories fails in the test (\textit{negative test}) and they are considered suspicious.
Based on the exclusion test, similarity check $S(\mathcal{T}_{i}, \mathcal{T}_{j})$ between the two trajectories can be defined as follows:
	  
	\begin{eqnarray} \nonumber
			S(\mathcal{T}_{i},  \mathcal{T}_{j})=
			\begin{cases}
			\frac{\abs{\mathcal{T}_{i} \cap  \mathcal{T}_{j}}}{Min\{\abs{\mathcal{T}_{i}},  \abs{\mathcal{T}_{j} }\}}  &  negative~test \\
			-1 & positive~test ,\\
			
			\end{cases}
			\label{Eq:4.12}
			\end{eqnarray}
		
		\begin{comment}
		
		\[
		(\mathcal{T}_1, \mathcal{T}_2) = \begin{dcases}
		\frac{| \mathcal{T}_1 \cap \mathcal{T}_2 |}
		{\min\{| \mathcal{T}_1 |, | \mathcal{T}_2 |\}} &  
		\textit{negative\> test}, \\
		-1 & \textit{positive\> test}
		\end{dcases}
		\]
		
\end{comment}
where minus one represents that $\mathcal{T}_{i}$ and $\mathcal{T}_{j}$ are distinct,
$\mathcal{T}_{i} \cap \mathcal{T}_{j}$ denotes the set of common RSUs found when
checking $\mathcal{T}_{i}$ and $\mathcal{T}_{j}$ using the check window and $|.|$
represents the length of a trajectory.

\begin{figure}[!t]
			\centering
			
			\includegraphics[width=1\linewidth]{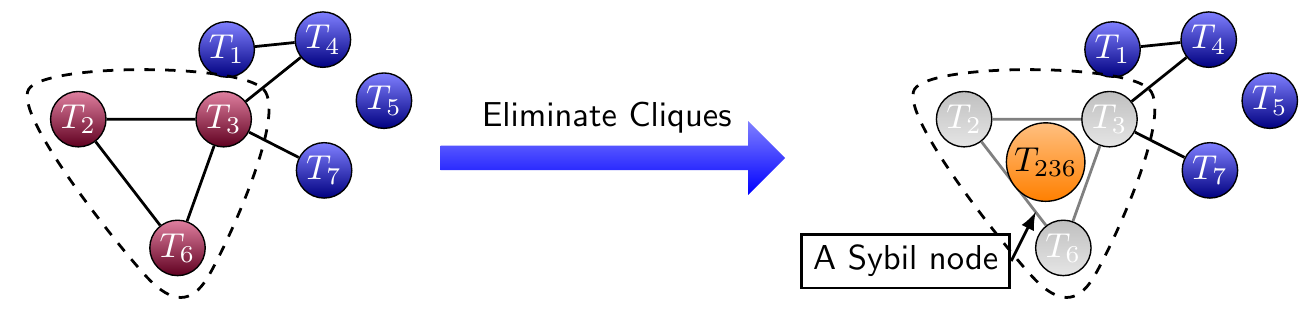}
			\vspace{0mm}
				\caption{The graph has one maximum clique, the triangle $\{\mathcal{T}_2,\mathcal{T}_3,\mathcal{T}_6\}$, and four more maximal cliques including the pairs $\{\mathcal{T}_1,\mathcal{T}_4\}$, $\{\mathcal{T}_3, \mathcal{T}_4\}$, and $\{\mathcal{T}_3, \mathcal{T}_7\}.$}
				\label{group}
		\end{figure}

\textbf{Phase 2: Building graph and eliminate cliques.} After conducting the similarity check phase, the Sybil detection problem can be transformed into finding all complete subgraghs (called cliques) in an undirected graph where graph vertices are the vehicles' submitted trajectories and edges represents the negative result of the exclusion test. We adopt maximum clique algorithm named "MCS" proposed by Tomita et. al. in~\cite{tomita2010simple} since it is a simple and faster branch-and-bound algorithm that can efficiently find a maximum clique for large number of graphs. As illustrated in Fig.~\ref{group}, the idea is to pick a maximum clique (fully connected graph) iteratively and to delete all vertices in that clique and all its corresponding edges from the graph till there are no more vertices left in the graph. In this way, a malicious vehicle is allowed to represent itself once to the event manger no matter how many forged trajectories it has generated.

	\section{Selection of PoW targets}
	\label{Selection of PoW targets}

In this section, we discuss how to select PoW target values that should be previously loaded in the look-up target table. The table maintains two values: the expected target value and the corresponding traverse time of a vehicle between two neighboring RSUs. First, we compute the target values experimentally. Then, we provide a mathematical model and compare it with the experimental results. Finally, we evaluate how the use of PoW in our scheme limits a malicious vehicle's ability to create multiple forged trajectories.
	\begin{comment}
		\begin{figure}[!t]
			\subfloat[Probability mass functions]{%
				\includegraphics[clip,width=3.3in]{Figures_POW/pmf_target_10_30_90_130}%
				\label{pmf}
			}
			
			\subfloat[Survival function]{%
				\includegraphics[clip,width=3.3in]{Figures_POW/cdf_target_10_30_90_130}%
				\label{survive}
			}
			\caption{Probability mass functions and survival functions of PoW target results with different traverse times.}
			\label{target results}
		\end{figure}
		\end{comment}
	
		\begin{figure}[!t]
			\captionsetup{justification=centering}
			\centering
			\includegraphics[width=\linewidth]{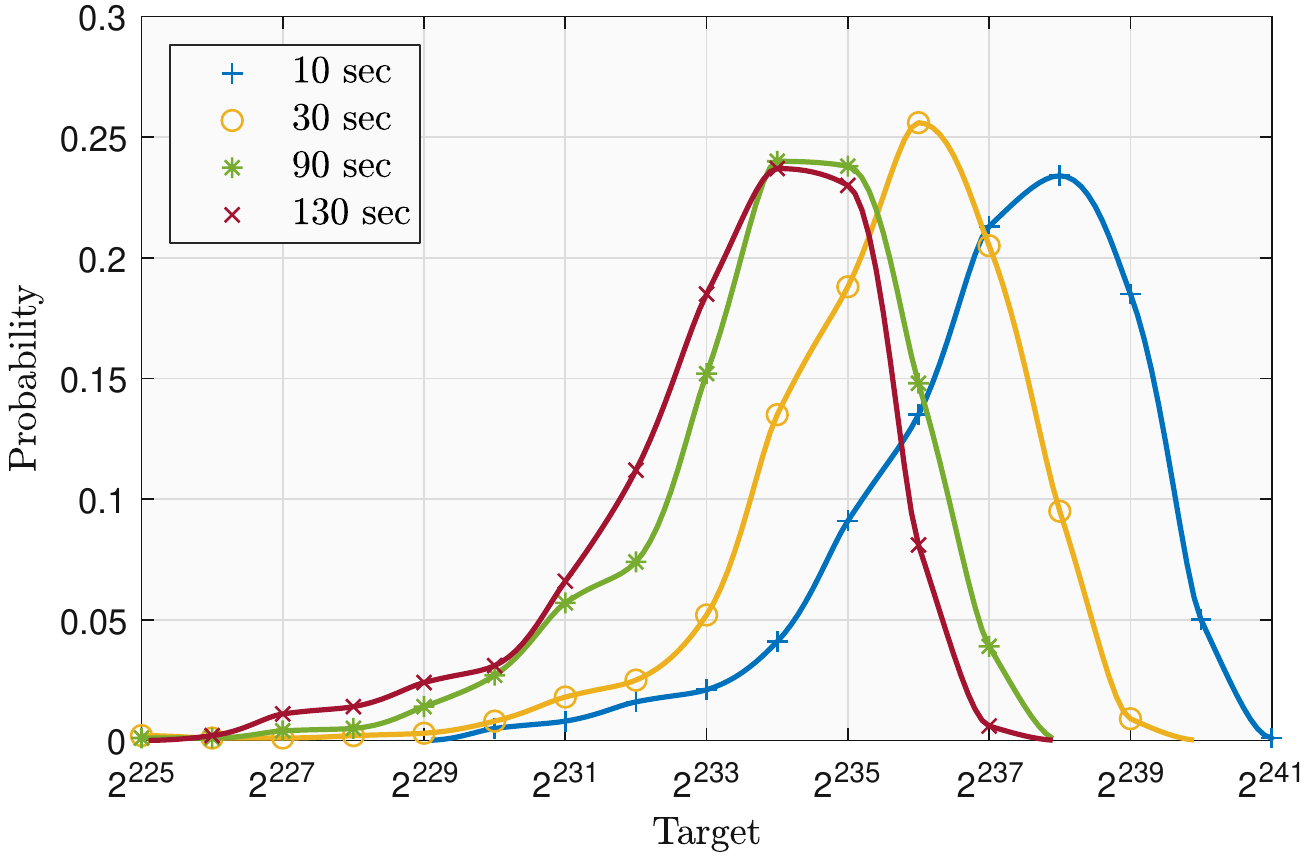}
			\caption{Probability mass function of computing PoW target values at different traverse times.}
			\vspace{0mm}
			\label{pmf}	
		\end{figure}
	
\begin{figure}[!t]
		\captionsetup{justification=centering}
		\centering
		\includegraphics[width=\linewidth]{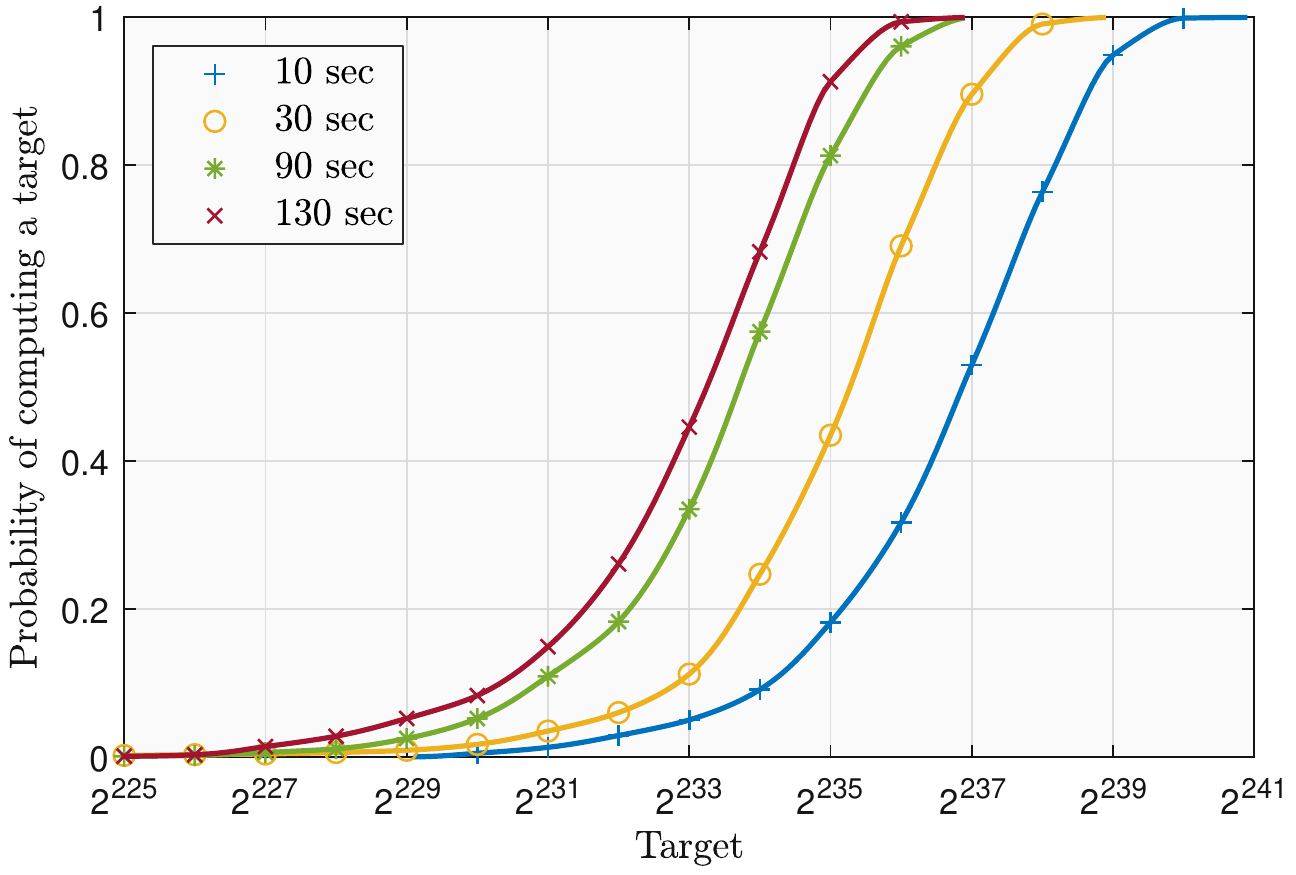}
		\caption{Cumulative distribution function of PoW target results with different traverse times.}
		\vspace{0mm}
		\label{survive}	
\end{figure}

\subsection{Experimental Results}
	
The steps to obtain the look up target table experimentally is as follows. First, using Raspberry Pi 3 devices, we compute the probability distribution function ($pmf$) of computing PoW target values by solving the PoW algorithm considering different values of traverse times, then we compute the cumulative distribution function (CDF) of solving the PoW target values. Finally, we used the CDF to obtain the look up target table. The details of our experiment is as follows;

to simulate a vehicle, we used Raspberry Pi 3 devices with 1.2 GHz Processor and 1 GB RAM. We selected four possible traverse times particularly 10, 30, 90 and 130 seconds. Note that these values represent the possible contact time between a moving vehicle and an RSU~\cite{kenney2011dedicated}. In addition, We ran the PoW algorithm 1000 times for each traverse time value. Then, we divided the obtained target values to equal ranges and then we estimated the probability of computing each of these target ranges.
%To illustrate, consider traverse time $t=10$ sec, the probability of getting a target $T=2^{239}$ is determined by the ratio of vehicles that was able to compute a target in a range of  $[2^{238}, 2^{239}]$.

Fig. \ref{pmf} shows the $pmf$ of the obtained target values at different traverse times. Using probability mass function in Fig. \ref{pmf}, the cumulative distribution function of computing a target is shown in Fig.~\ref{survive}. Note that the CDF is the probability that a random variable ($T$, that is the target value in our case) takes a value less than $t$~\cite{lee2003statistical} as follows:
% $$S(x)=Pr[X>x]$$
 
 $$Pr[T<t]= \int_{-\infty}^{t}pmf(t) \: dt$$

For instance, from Fig.~\ref{survive}, at traverse time $t=130 sec$, the probability of computing a target lower than $T=2^{239}$ is $0.98$. That is, $98\%$ of vehicles are able to compute a target value equal or greater  than $T=2^{239}$ at traverse time equal to $130 sec$.

	\begin{comment}		

	\begin{figure}[!h]
		\captionsetup{justification=centering}
		\centering
		\includegraphics[width=\linewidth]{Figures_POW/target_over_time_succ_rate_95_90_85_80}
		\caption{Experimentally determined target values at different traverse times.}
		\vspace{0mm}
		\label{fig:target_over_time_succ_rate_95_90_85_80}
		
	\end{figure}
	
	\end{comment}
	
		\begin{figure}[!t]
		
			\centering
			\includegraphics[width=\linewidth]{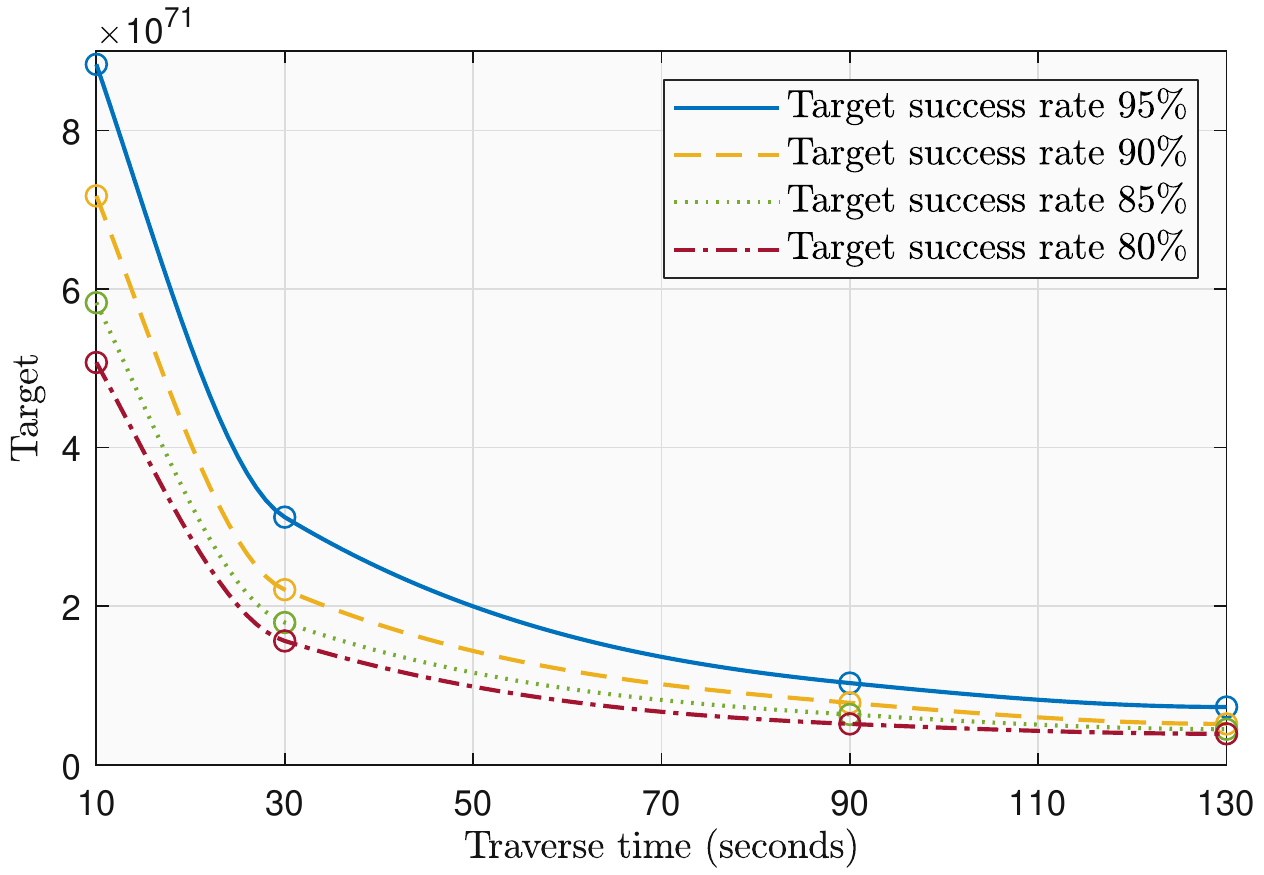}
			\caption{Experimentally determined target values at different traverse times (target lookup table).}
			\vspace{0mm}
			\label{fig:target_over_time_succ_rate_95_90_85_80}
			
		\end{figure}	
		
 %shows the expected target values at different target success rates of vehicles that were able to obtain a certain target value.
	
Next, using the results shown in Fig.~\ref{survive}, we have estimated the expected targets at different traverse times as illustrated in Fig.~\ref{fig:target_over_time_succ_rate_95_90_85_80} at different target success rate values. To illustrate, 95\% target success rate indicates that $95\%$ of vehicles are able to compute a certain target value. For example, at traverse time 50 seconds, 95\% of vehicles are able to solve the PoW below a target of $2 \times 10^{71}$ while 80\% of vehicles were able to compute a target below $1 \times 10^{71}$. As depicted in Fig.~\ref{fig:target_over_time_succ_rate_95_90_85_80}, as the traverse time of the vehicle increases, a lower target value is expected by the next RSU. This is because the vehicles have more time to solve the PoW algorithm. Now, the lookup table can be constructed and loaded to each RSU so that RSUs can use that table to verify whether a vehicle successfully solved a PoW puzzle. 
	
	 %In addition at a certain traverse time, the higher the target value, the higher the percentage of vehicle who are able to get that target. At the end,
	
\subsection{Mathematical Model}

As discussed above, the experimental method can be used to determine the target look-up table. However, experimental method is complex due to many parameters involved such as the hardware used, the hash function used in the PoW algorithm, and the traverse times. Therefore, in this subsection, we aim to provide a mathematical model for the probability of computing targets and compare the model with our experimental results.

%compute the target value, our experiment can be described as determining the probability of successes in randomly drawing $n$ times from a population of size $N$, where $n$ is the number of hashes per sample and $N$ is the output range of the hash function. Within $N$, there are $K$ PoW puzzle solutions that can satisfy certain required targets, and since any hash result below a certain target is a correct solution, $K$ is the target value. Henceforth, we can be map our experiment to a hypergeometric distribution. The probability mass function of the hypergeometric distribution is given by~\cite{berkopec2007hyperquick}: 
	\begin{figure}[!t]

			\centering
			\includegraphics[width=\linewidth]{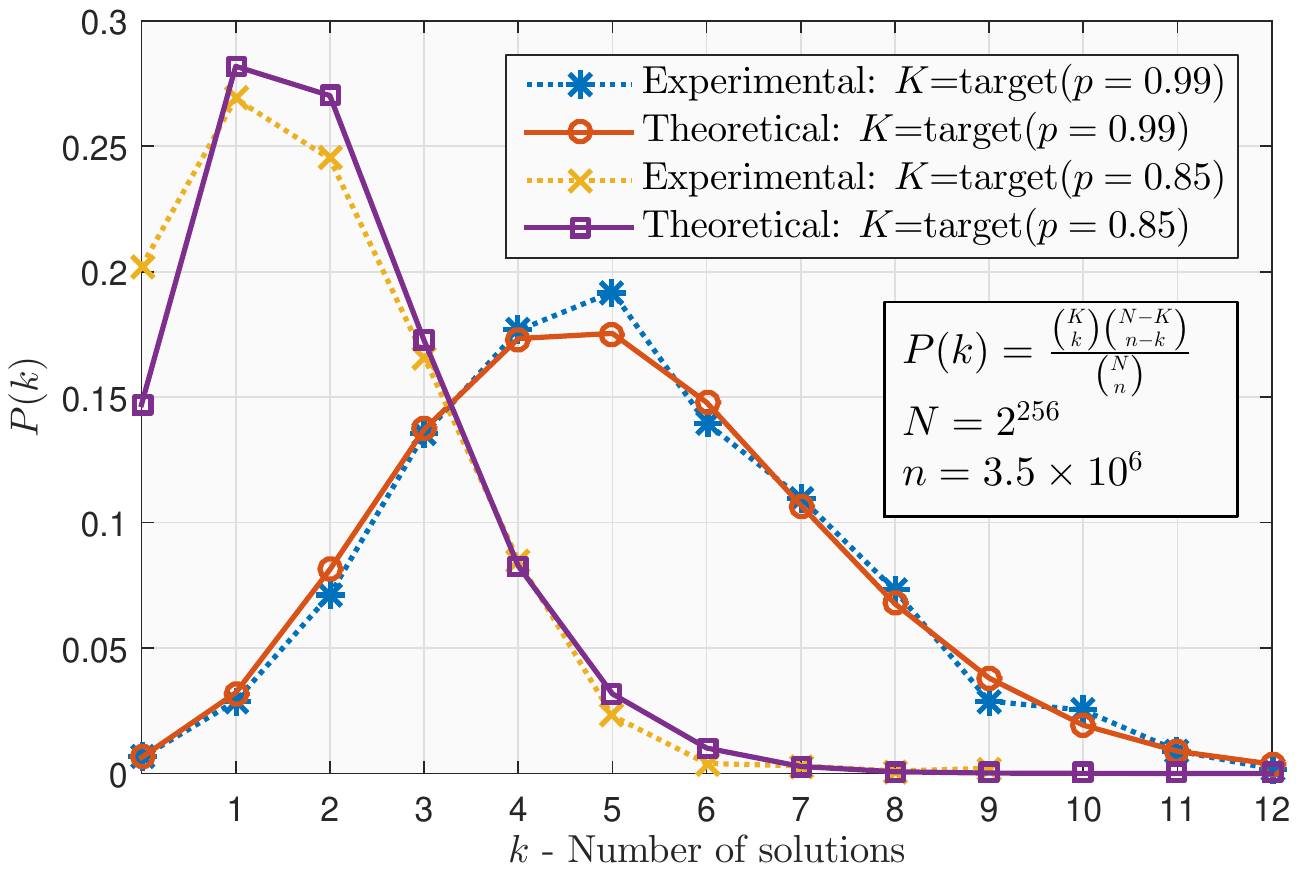}
			\caption{Comparison between the experimental and theoretical probability distribution of finding $k$ PoW solutions within $t=90 sec.$ for $K=16.74\times10^{70}$ and $K=6.34\times10^{70}$ at target success rate 99 \% and 85 \% respectively.}
			\vspace{0mm}
			\label{fig:experiment_vs_hypgemdist}
			
		\end{figure}
Successfully solving $k$ PoW puzzles by a vehicle during a specific traverse time between two RSUs can be modeled by a hypergeometric distribution \cite{berkopec2007hyperquick} model. This model describes the probability of having $k$ successes in $n$ draws from a finite pool of size $N$ that contains $K$ objects that are successful. This model can be mapped to our PoW computation as follows: a vehicle' hash rate during the traverse time is equivalent to the number of $n$ draws from a pool of hashes of size $N$ to get $k$ successful solutions that satisfy the target which is $K$ in that case. Therefore, we are looking for a probability of success in solving $k$ PoW puzzles using the following formula:

	\begin{equation}
	\label{eq:1}
	P(k)= \frac{{K \choose k}{N-K \choose n-k}} {{N \choose n}}
	\end{equation}
	
	We consider the following parameters:
	\begin{itemize}
		\item $N = 2^{256}$ is the output space of the hash function SHA-256.
		\item $K$ is the target value  with a certain success rate at a particular traverse time. For the value of $K$, we used results obtained in Fig.~\ref{fig:target_over_time_succ_rate_95_90_85_80} to get the target values at the traverse time equal to $90$ sec. Basically, we used two values, $16.73 \times 10^{70}$ and $6.34 \times 10^{70}$ which are the target values at target success rate 99\% and 85\% respectively.
		\item $n=3.5\times10^6$ is the number of hashes per traverse time $t=90 sec$. This value is estimated in python.
	\end{itemize}
	Using the aforementioned parameters, we are able to obtain the probability of finding $k$ PoW solutions within a certain traverse time $t$~\cite{RefWorks:104}. However, due to large numbers in our case, it can be hard to compute the probability mass function using Eq.~\ref{eq:1}. Nevertheless, since the range of $k$ is small, we can approximate the hypergeometric distribution using the Poisson distribution according to~\cite{zelterman1999models} as follows:
	\begin{equation}
	\label{eq:approximate}
	F(k)= \frac{\lambda^k \exp^{-\lambda}} {k!}
	\end{equation}
	where $\lambda=\frac{nK} {N}$. 
	
Both the experimental and theoretical results are shown in Fig.~\ref{fig:experiment_vs_hypgemdist}. As per Fig.~\ref{fig:experiment_vs_hypgemdist}, We can conclude that, our experimental results match the approximated hypergeometric distribution. In addition, by using Eq.~\ref{eq:approximate}, we can generate the target look up table in Fig.~\ref{fig:target_over_time_succ_rate_95_90_85_80} mathematically by merely knowing the hash rate during a certain traverse time.

		\subsection{Importance of using PoW}
%The formula used in Eq.~\ref{eq:1} is used to determine the probability of obtaining $k$ PoW solutions at a certain traverse time. 

%. Assuming a vehicle traverse along $l$ RSU as $\{R_1, R_2,\cdots,R_l\}$. 

Now, we examine the effect of PoW algorithm for honest vehicles and malicious vehicles. As discussed earlier, benign vehicles should submit one valid PoW solution for every puzzle. However, in case of a malicious vehicle trying to get $i$ trajectories, each of length $j$, it should be able to solve the PoW puzzle at least $i$ times for every RSU it encounters. Failing to solve one puzzle at any RSU leads in failing to create one of the trajectories. The probability of creating $i$ trajectories of length $j$ RSUs can be obtained as follows: First, the probability of computing $i$ valid PoW puzzle solutions at an RSU can be obtained by applying the survival function to the formula in Eq.~\ref{eq:approximate}, where, the survival function means the probability that a random variable ($X$) takes a value greater than $i$ as follows,
\begin{equation}
 S(i)=Pr(X\geq i)=\sum_{k=i}^{\infinity}F(k)
 	\end{equation}
 Then, the overall successful probability of creating $i$ trajectories of length $j$ is computed as $Pr(X \geq i)^{j}$ and can be expressed mathematically  as follows:

	  	\begin{equation}
	\label{max}
 Pr(X \geq i)^{j}= \left(\sum_{k=i}^{\infinity} F(k)
 \right)^j 	\approx \left(\sum_{k=i}^{\infinity} \frac{\lambda^k \exp^{-\lambda}} {k!}\right)^j
	\end{equation}
	  
	  	\begin{comment}
	  \begin{equation}
	\label{max}
	P(\mathcal{T}^j_i)=  \left(S(i)\right)^j = \left(\sum_{k=i}^{k =\infinity} P(k)
\right)^j
	\end{equation}
	  	\end{comment}

\begin{comment}

	\begin{equation}
	
	P(\mathcal{T}^j_i)= \left(\sum_{\bar{k}=i}^{\bar{k} =\infinity}
	\frac{\binom{K}{\bar{k}}\binom{N-K}{n-\bar{k}}}{\binom{N}{n}}\right)^j
\label{max}
	\end{equation}
		
		\end{comment}

In order to evaluate the importance of using the PoW, we define the following two key metrics:

\begin{itemize}
		\item \textit{One trajectory computation successful probability ($P_O$)}: is defined as the probability that an honest vehicle is able to compute one valid trajectory. This probability  can be mathematically computed by using $i=1$ in Eq.~\ref{max} since to be able to create at least one trajectory, an honest vehicle should be able to compute the target at each RSU it encounters.  
\item \textit{Multiple trajectories computation successful probability ($P_M$):} is defined as the probability that a malicious vehicle can compute multiple trajectories i.e., solving multiple puzzles and compute the expected targets at each RSU it encounters. This probability can be obtained by using $i > 1$ in Eq.~\ref{max}
since to create multiple trajectories, the malicious vehicle should be able to solve more than one  puzzle at each RSU it encounters. %If a malicious vehicle fails to compute the targets of at one RSU, the vehicle can not obtain a valid trajectory associated with that target.
	\end{itemize}

		\begin{figure}[!t]
		
			\centering
			\includegraphics[width=\linewidth]{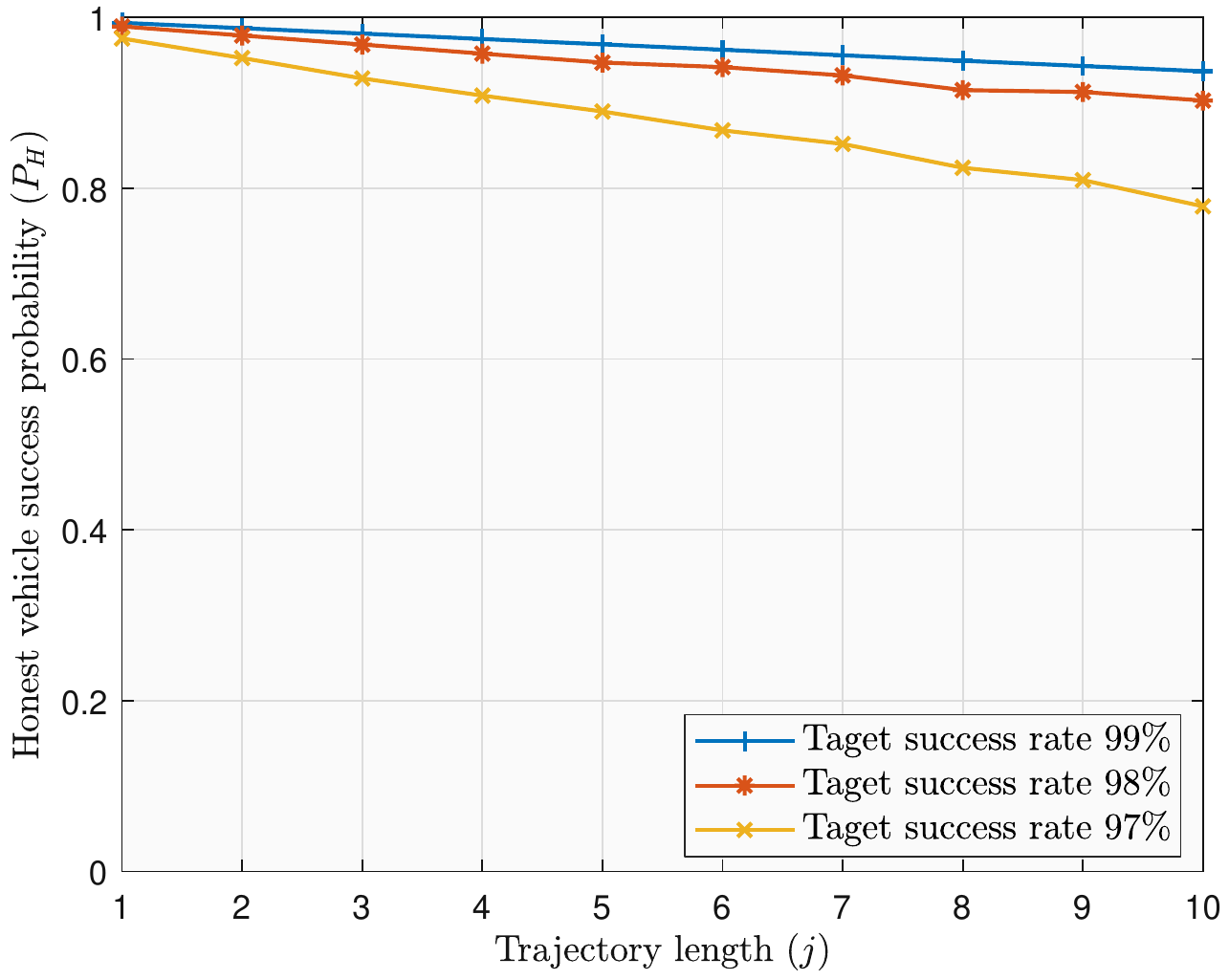}
			\caption{One trajectory computation successful probability in generating a valid trajectory as a function of trajectory length.}
			\vspace{0mm}
			\label{fig:honest_success_probability}
			
		\end{figure}
	
\begin{figure*}[!t]
	\setlength{\abovecaptionskip}{0.1cm}
	\setlength{\belowcaptionskip}{-0.7cm}
	\centering
	\subfloat[Target success rate 98\%.\label{98}]
    {\includegraphics[width=0.45\linewidth]{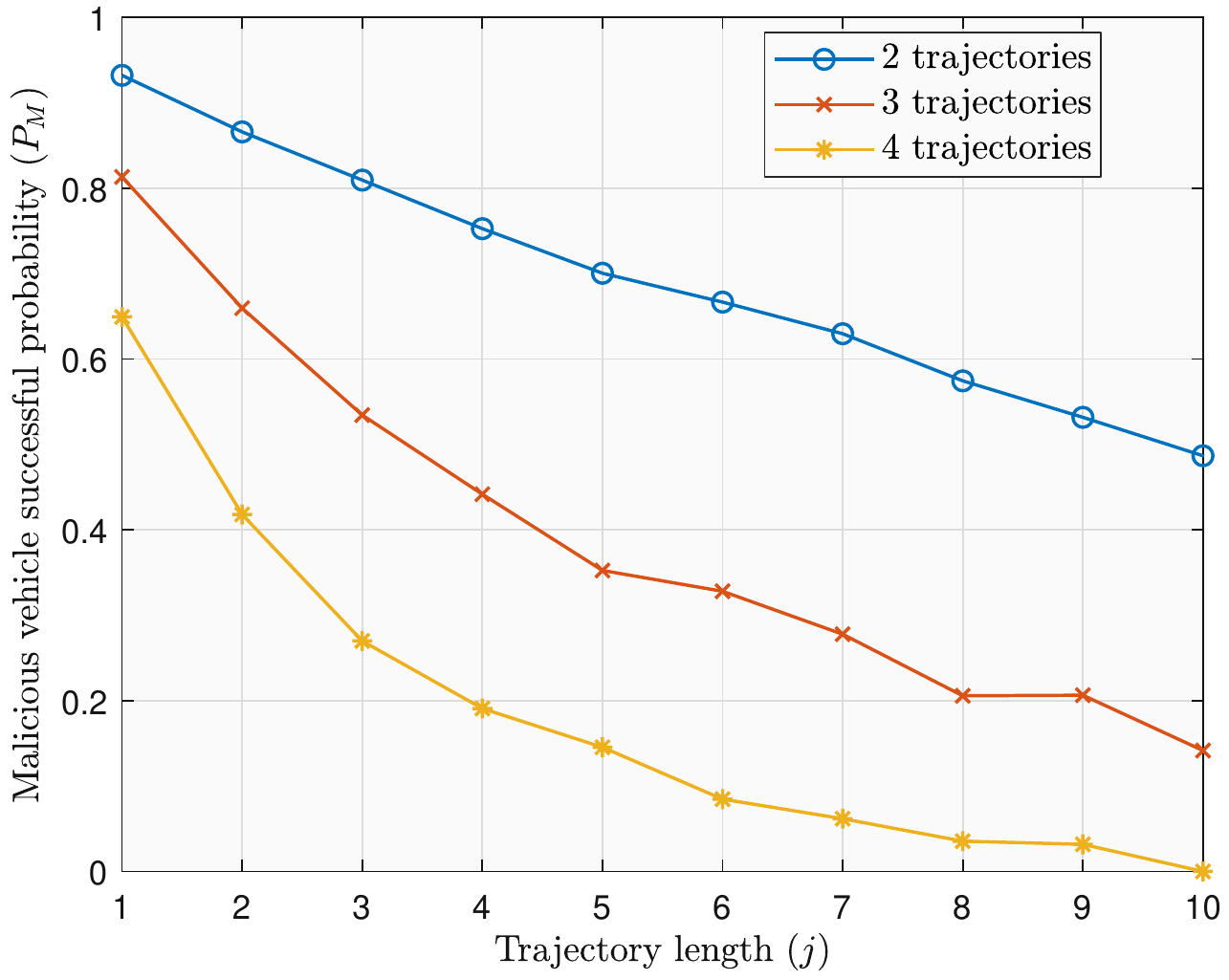}}
    \subfloat[Target success rate 99\%.\label{99}]
	{\includegraphics[width=0.45\linewidth]{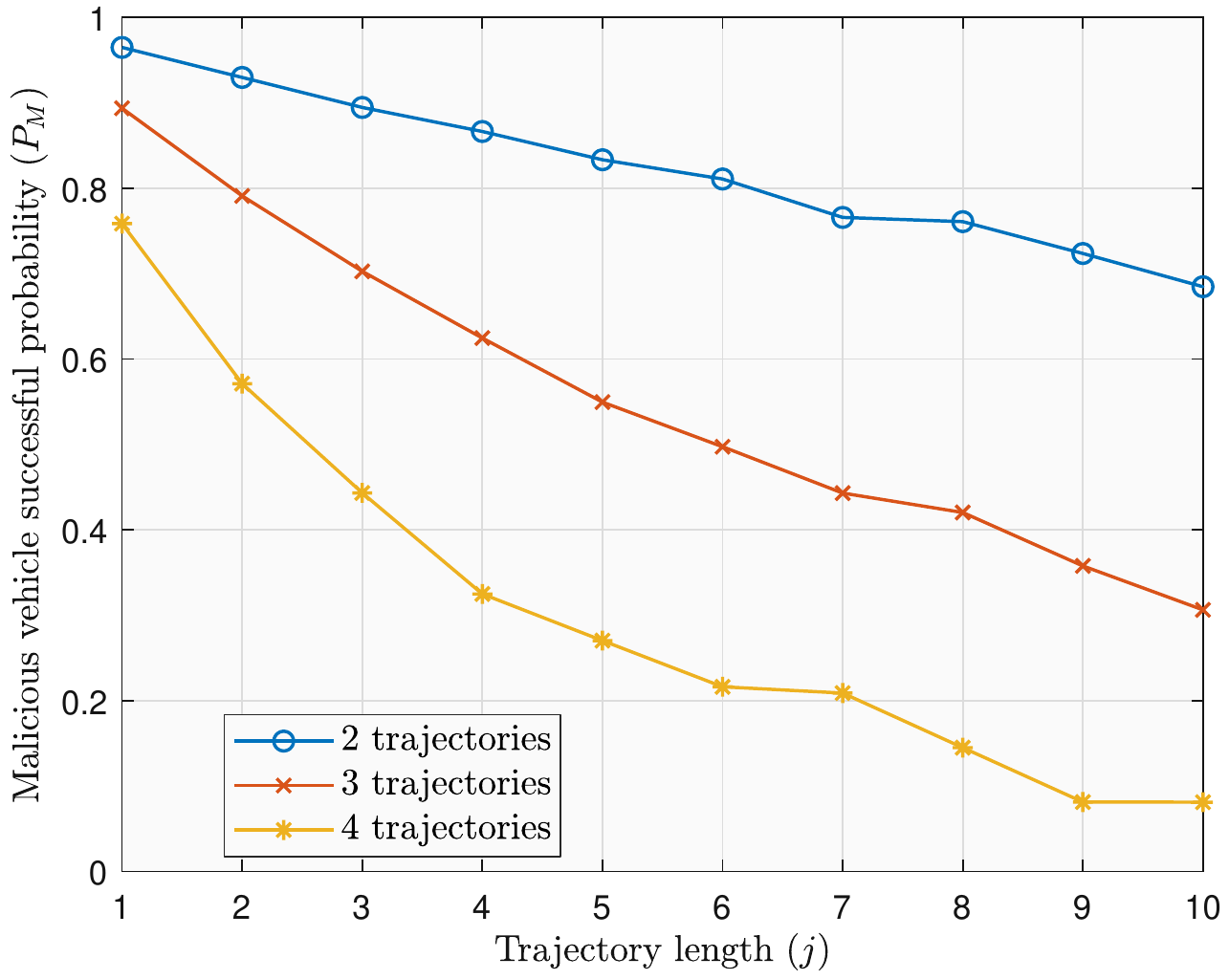}}

	\caption{Malicious vehicle success probability in generating multiple trajectories with increasing trajectory length considering 98\% and 99\% success rates of solving a puzzle.}
	\label{Malicious vehicle success rate}
	
\end{figure*}

 Considering the target values obtained in Fig.~\ref{fig:target_over_time_succ_rate_95_90_85_80} at traverse time equal to $90$ sec., Fig.~\ref{fig:honest_success_probability} shows the one trajectory computation successful probability ($P_O$) in generating a trajectory with different trajectory lengths as well as at different target success rates namely, 99\%, 98\% and 97\%. As shown in Fig.~\ref{fig:honest_success_probability}, The higher the target success rate, the higher the probability of honest vehicle is able to compute a valid longer trajectory. However, setting the expected target too low at certain traverse time results in that honest vehicle will slightly sacrifice the possibility of computing a valid trajectory. Fig.~\ref{Malicious vehicle success rate} shows the probability of creating $2, 3$ and $4$ trajectories with a trajectory length up to 10 as well as at different target success rates namely, 99\% and 98\%. It is clearly seen from Fig.~\ref{Malicious vehicle success rate} that the probability of creating multiple trajectories decreases significantly with the increase of trajectory length which can reduce the success of Sybil attacks by making creating multiple trajectories difficult. It can be concluded that:

\begin{enumerate}[label=(\roman*)]

	\item Increasing the difficulty level of PoW puzzle by lowering value significantly reduces the multiple trajectories computation successful probability ($P_M$). However, setting the target value too low can result in honest vehicles not being able to compute a successful trajectory. Therefore, a target value should be well chosen to enable honest vehicles to create trajectories with high probability while reducing the chance of the malicious vehicles to create many trajectories. 
	
	\item As the trajectory length increases, the lower the multiple trajectories computation successful probability and it becomes difficult for an malicious vehicle to compute multiple trajectories.

	%The trajectory length  is an important factor to detect the Sybil attack. Selecting lower  trajectory length limit values gives a malicious vehicle more chance to create multiple trajectories by solving multiple PoW puzzles. Therefore, trajectory length limit should be well chosen to achieve high detection rates for Sybil attacks.
\end{enumerate}
	\begin{comment}
	\begin{figure}[!t]
		\captionsetup{justification=centering}
		\centering
		\includegraphics[width=\linewidth]{Figures_POW/honest_success_probability}
		\caption{Successful probability for honest vehicles in generating a valid trajectory with increasing trajectory length.}
		\vspace{0mm}
		\label{fig:honest_success_probability}
		
	\end{figure}
	\end{comment}

	\begin{comment}
	\begin{figure}[!h]
		\subfloat[ P = 0.98\%]{%
			\includegraphics[clip,width=3.3in]{Figures_POW/malicioussuccessprobability98}%
			\label{98}
		}
		
		\subfloat[P = 0.99\%.]{%
			\includegraphics[clip,width=3.3in]{Figures_POW/malicious_success_probability99}%
			\label{99}
		}
		\caption{Malicious vehicle success probability in generating multiple trajectories with increasing trajectory length considering 98\% and 99\% success rates of solving a puzzle.}
		\label{Malicious vehicle success rate}
	\end{figure}
	
	\end{comment}

	\begin{comment}

	\begin{figure}[!h]
		\subfloat[Target success rate 98\%.]{%
			\includegraphics[clip,width=3.3in]{Figs/malcious}%
			\label{98}
		}
		
		\subfloat[Target success rate 99\%.]{%
			\includegraphics[clip,width=3.3in]{Figs/malcious1}%
			\label{99}
		}
		\caption{Malicious vehicle success probability in generating multiple trajectories with increasing trajectory length considering 98\% and 99\% success rates of solving a puzzle.}
		\label{Malicious vehicle success rate}
	\end{figure}
	
		\end{comment}
	
		\begin{figure}[!t]
			\captionsetup{justification=centering}
			\centering
			\includegraphics[width=\linewidth]{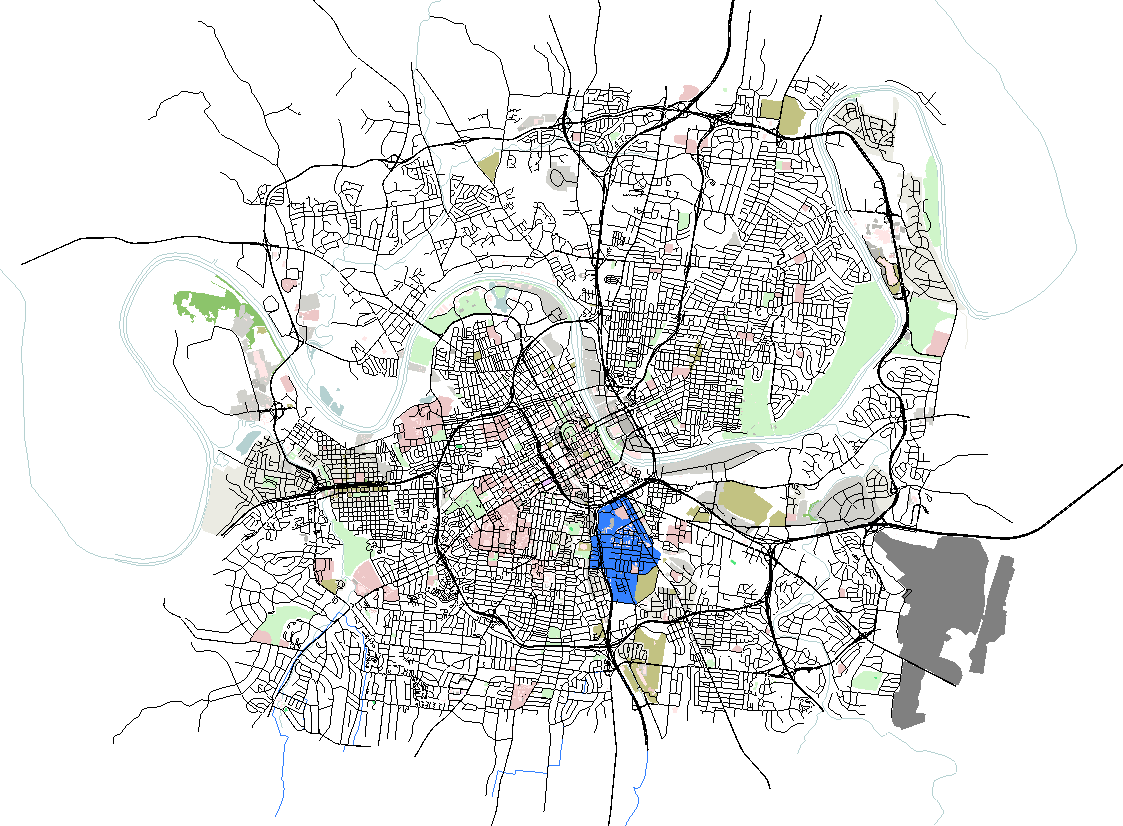}
			\caption{Map of Nashville city, TN, USA used for simulation.}
			\vspace{0mm}
			\label{fig:nashville_map}
			
		\end{figure}

		\begin{figure*}[!t]
	\setlength{\abovecaptionskip}{0.1cm}
	\setlength{\belowcaptionskip}{-0.7cm}
	\centering
	
	\subfloat[Check window size versus false positive rate.\label{fig:E2ECT}]
	{\includegraphics[width=0.3333333333\linewidth]{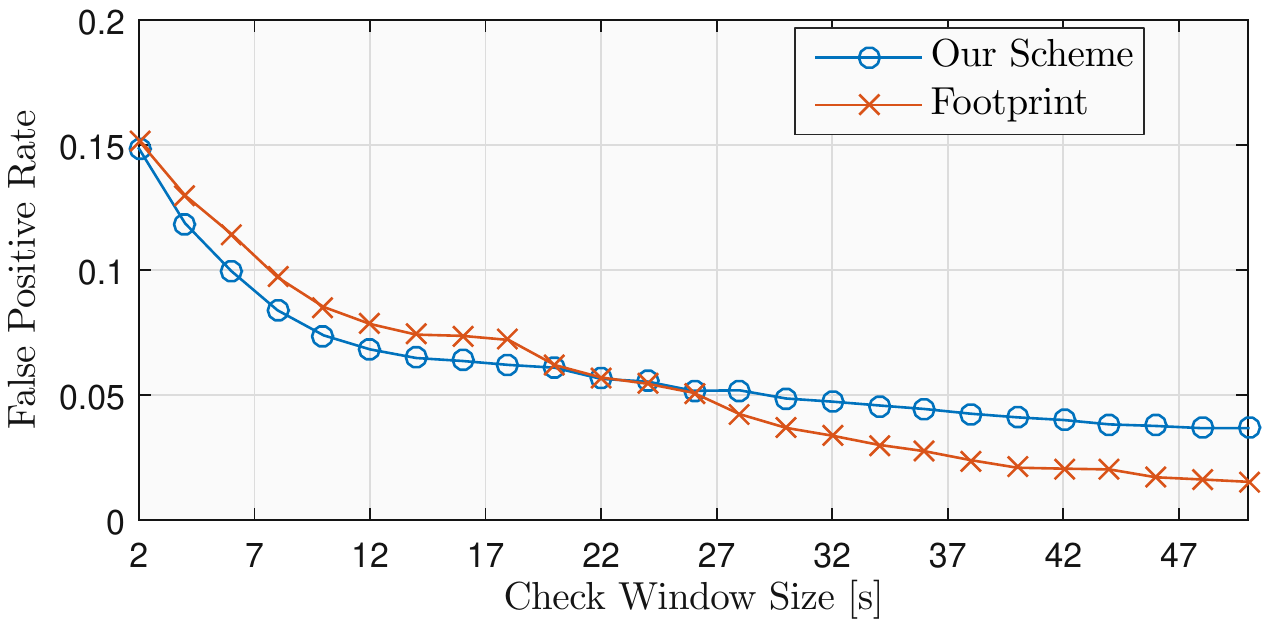}}	
	\subfloat[Check window size versus false negative rate.\label{fig:E2ETP}]
	{\includegraphics[width=0.3333333333\linewidth]{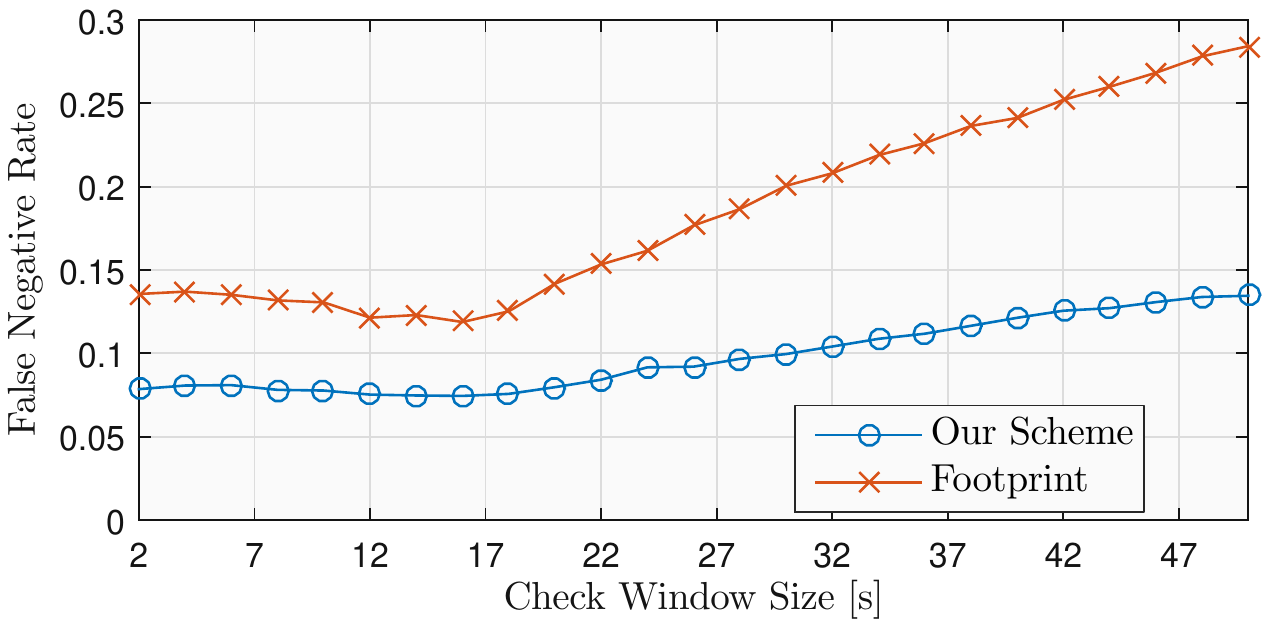}}	
	\subfloat[Check window size versus detection rate.\label{fig:E2EPDR}]
	{\includegraphics[width=0.3333333333\linewidth]{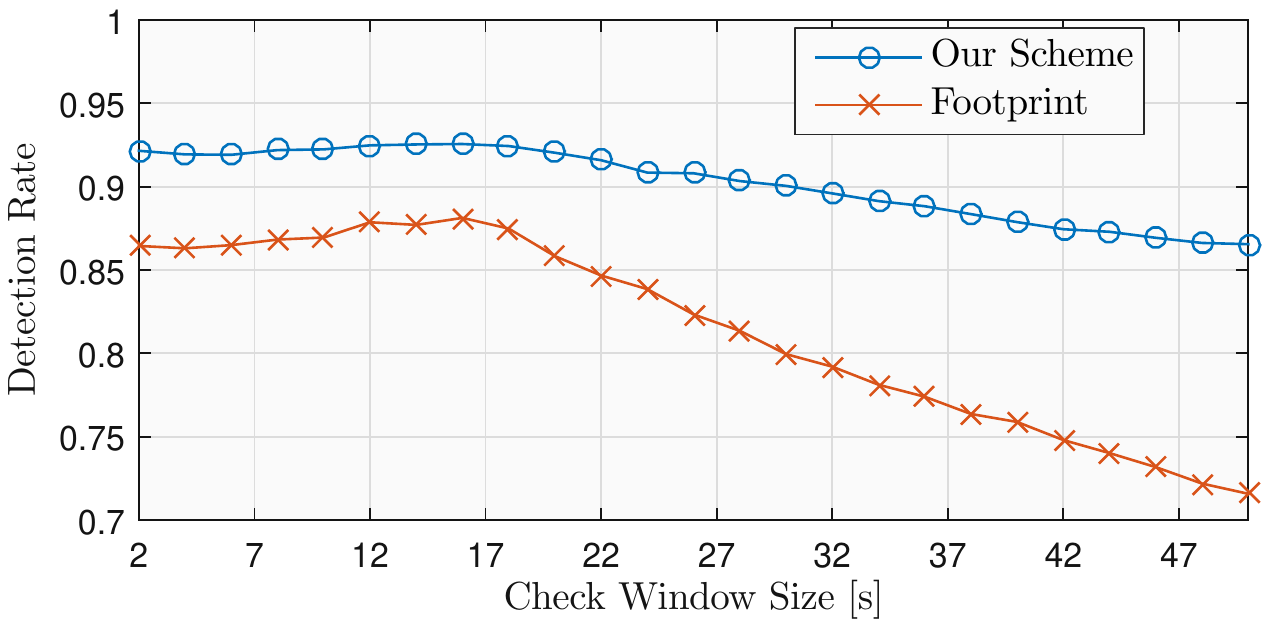}}\\
	\vspace{-3 mm}
	\subfloat[Trajectory length limit versus false positive rate.\label{fig:HbHCT}]
	{\includegraphics[width=0.3333333333\linewidth]{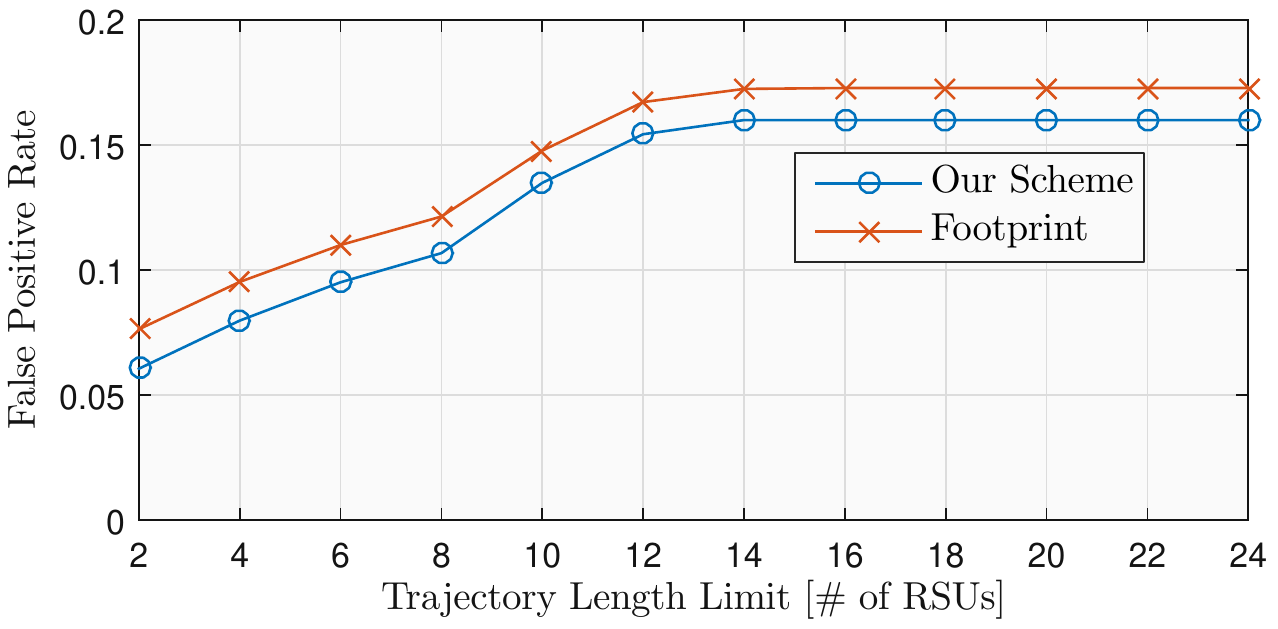}}	
	\subfloat[Trajectory length limit versus false negative rate.\label{fig:HbHTP}]
	{\includegraphics[width=0.3333333333\linewidth]{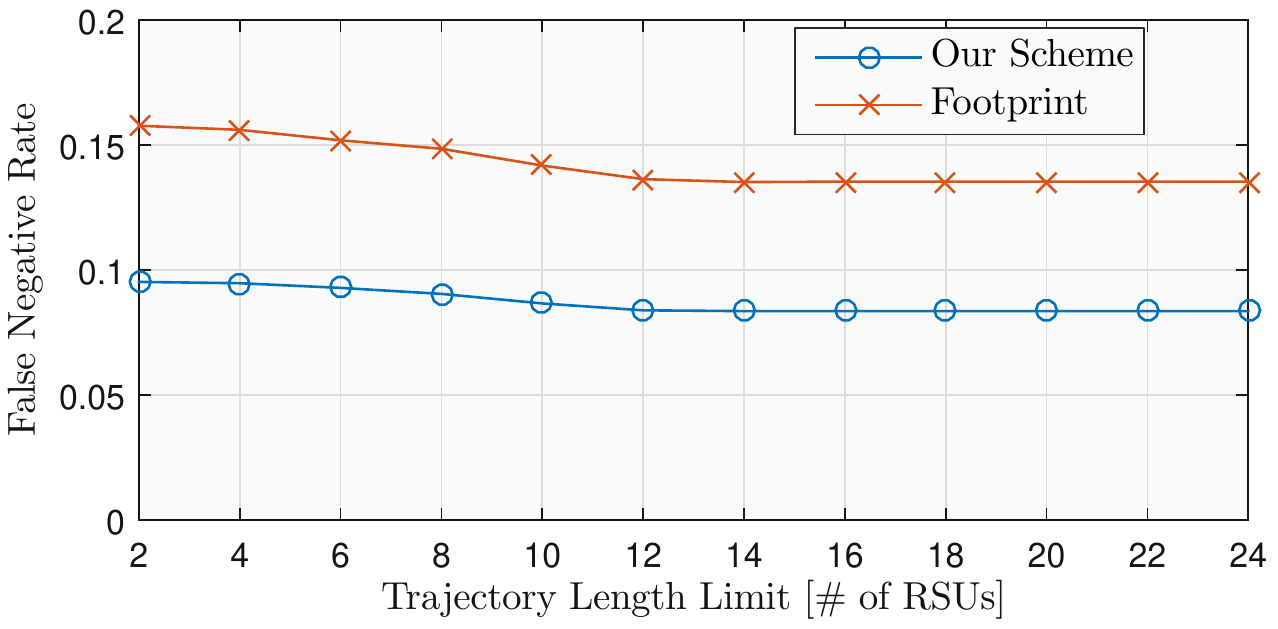}}	
	\subfloat[Simulation results at variable trajectory length limit.\label{fig:HbHPDR}]
	{\includegraphics[width=0.3333333333\linewidth]{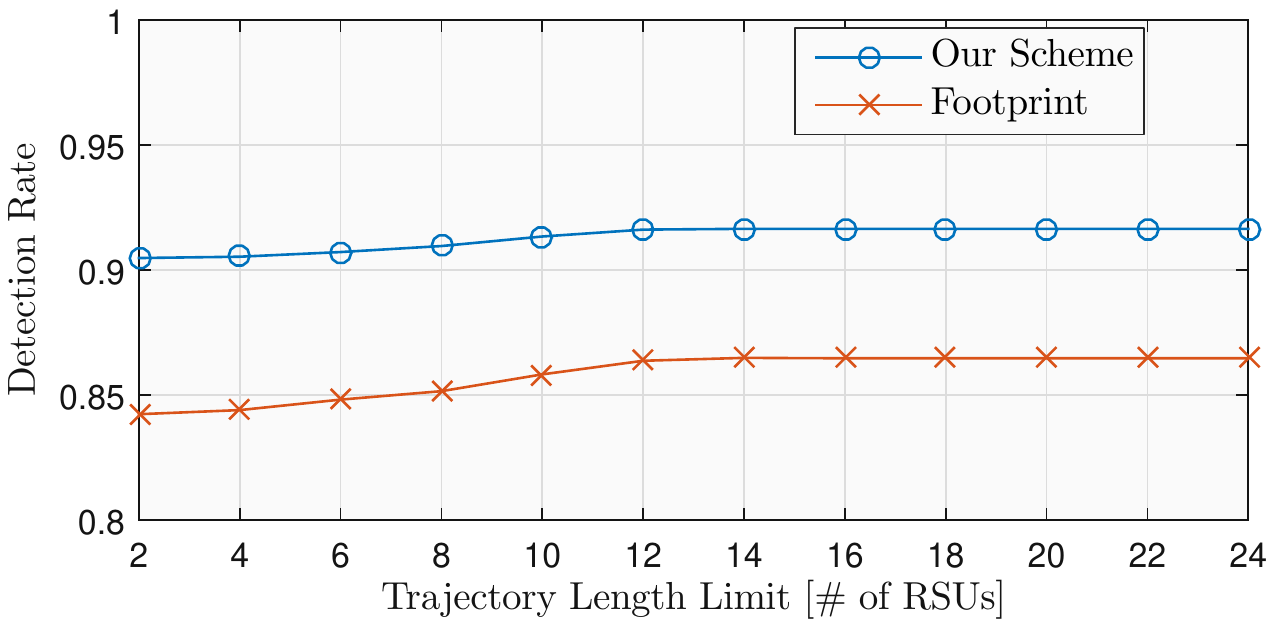}}	
	\caption{Simulation results comparison at variable check window sizes and trajectory length limit.}
	\label{fig:ns3_new}
	
\end{figure*}
			
\begin{figure*}[!t]
	\setlength{\abovecaptionskip}{0.1cm}
	\setlength{\belowcaptionskip}{-0.7cm}
	\centering
	\subfloat[Check window size versus detection time.\label{checkcomptime}]
    {\includegraphics[width=0.3333333333\linewidth]{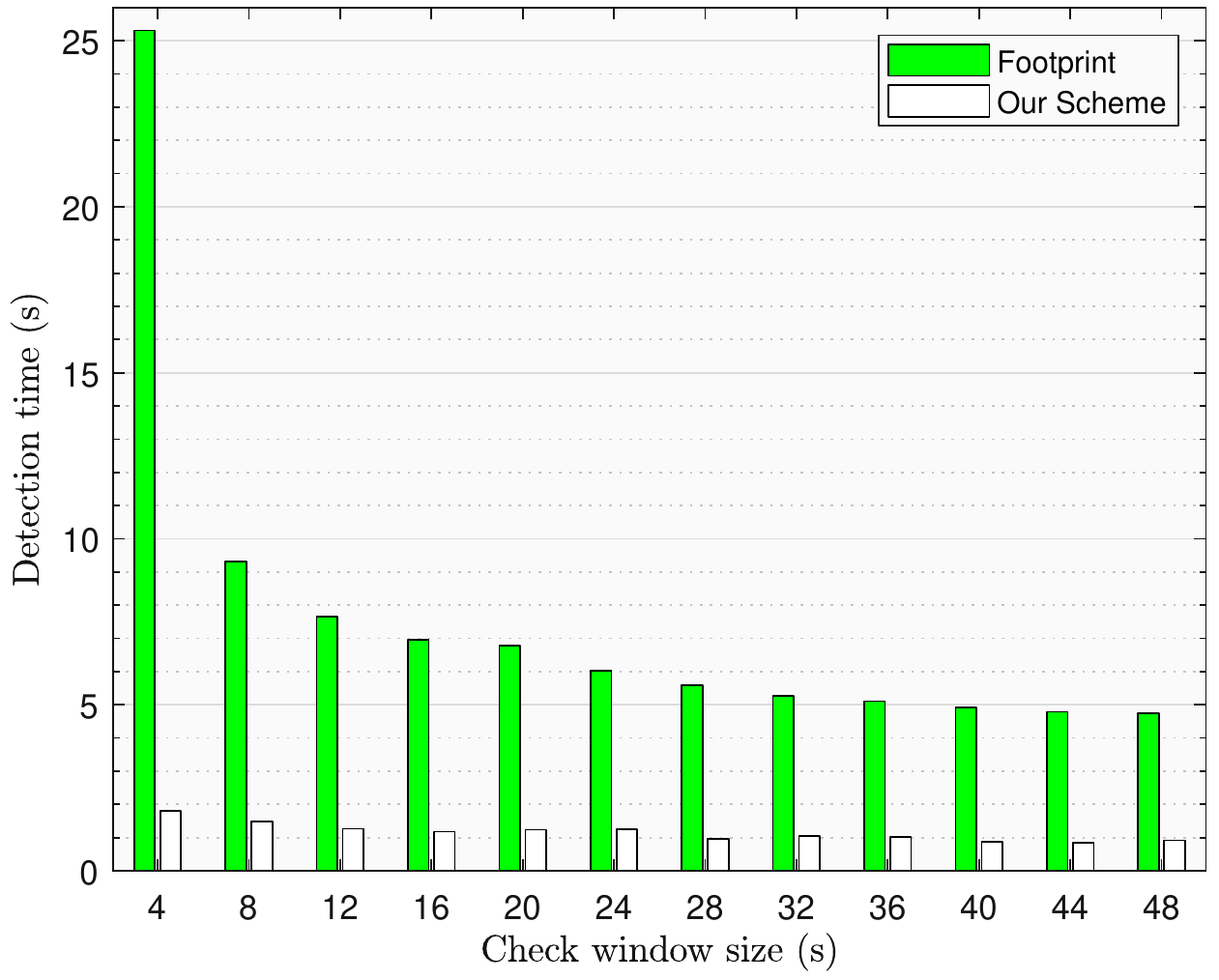}}
    \subfloat[Trajectory length limit versus detection time.\label{trajcomptime}]
	{\includegraphics[width=0.3333333333\linewidth]{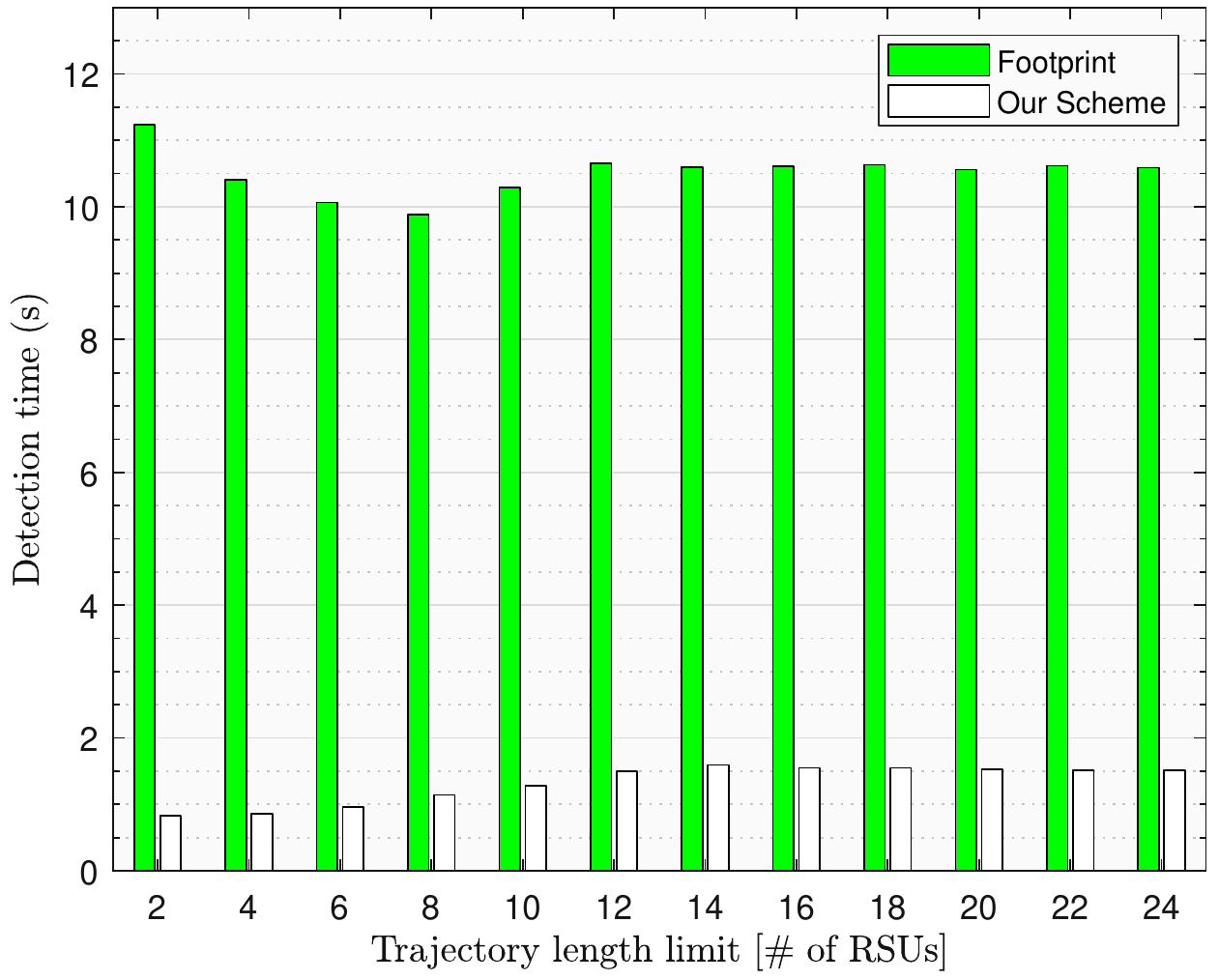}}
    \subfloat[Trajectory length limit versus detection time. \label{forgecomptime}]
	{\includegraphics[width=0.3333333333\linewidth]{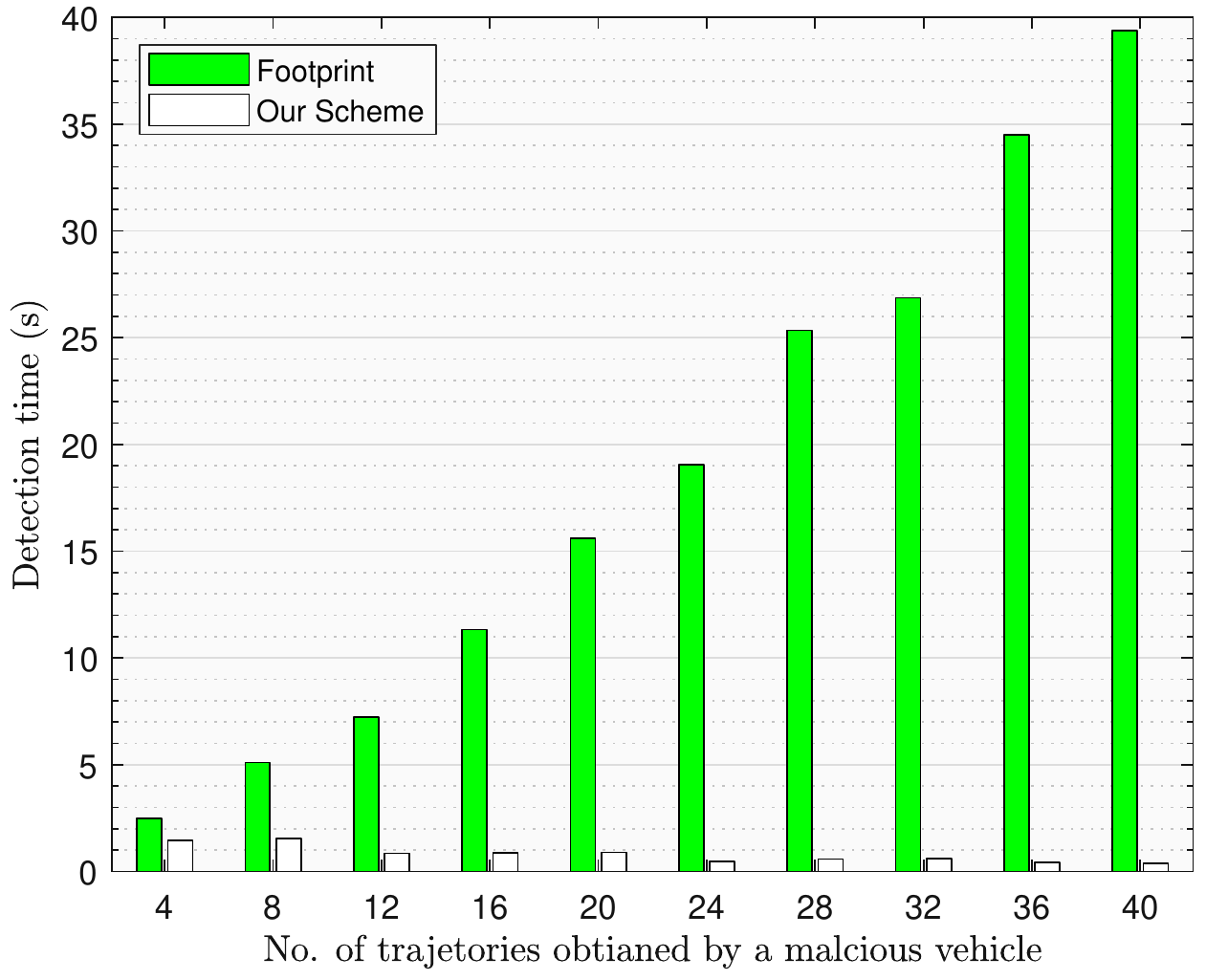}}
	
	\caption{Detection time comparison.}
	\label{fig:computation_new}

\end{figure*}
		
\section{Security and Evaluation Analysis}
	\label{performance}
	In this section, we evaluate our proposed scheme and compare it with the Footprint. We first start with performance evaluation of Sybil attack detection and show by simulations the efficiency of our scheme in terms of detecting Sybil attacks and the detection time. Then, we discuss the security and privacy analysis of our scheme.  
	
	\subsection{Sybil Attack Detection}
In this subsection, we discuss the performance of our proposed scheme by investigating different system parameters in recognizing forged trajectories (provided by a malicious vehicle) and actual ones (issued by honest vehicles).
	\subsubsection{Simulation Design}
	\label{ExperimentSetup}
  Our proposed scheme was implemented using Python, and trajectories were generated based on real roadmap data. First, using OpenStreetMap project \cite{OpenStreetMap}, we extracted the roadmap shown in Fig. \ref{fig:nashville_map} for Nashville city, TN, USA. The dimensions of the map are 75.5 km $\times$ 33 km. Second, using SUMO \cite{SUMO2012}, we generated random routes for about 160 vehicles.  RSUs were deployed at the edges of routes produced by the SUMO. Then, we recorded the actual trajectories generated by vehicles and truncated each trajectory into subtrajectories of lengths varying between 10 and 15 RSU's. We then assigned every trajectory with a random start time within a 5-second window and a traverse time between two RSUs, chosen at random between 10 and 130 seconds. Accordingly, our experiment simulates a large number of vehicles with random trajectories, all starting their route within a 5 seconds time window, and each travelling with random but constant speed. From all vehicles in our simulation, we arbitrarily chose 10 per cent to simulate malicious vehicles. For every malicious vehicle, we randomly created between 1 and 10 forged Sybil trajectories. 
	
	%\begin{figure}[tbp]
	%\setlength{\abovecaptionskip}{0.3cm}   % 0.5cm as an example
	%\setlength{\belowcaptionskip}{0.0cm}   % 0.5cm as an example
	%\centering
	%\includegraphics[scale=0.5]{Figures_Simu/nashville_map.png}
	%\caption{Map of Nashville, TN used for simulation.}
	%\vspace{0mm}
	%\label{fig:nashville_map}
	%\end{figure}
	
	\subsubsection{Considered key metrics}
	We first define the following terms regarding the classifications of output of running the detection algorithm: 
	\begin{itemize}
		\item \textit{False positives} ($FP$): Number of Actual trajectories that are incorrectly considered Sybil trajectories .
		\item \textit{True negatives} ($TN$): Number of actual trajectories that are correctly considered actual trajectories. 
		\item \textit{True positives} ($TP$): Number of Sybil trajectories that correctly considered Sybil trajectories.
		\item \textit{False negatives} ($FN$): Number of Sybil trajectories that are incorrectly considered actual trajectories.
	\end{itemize}

	Based on the previous classifications, we consider the following three key metrics in our performance evaluation:
	\begin{enumerate}
		\item \textbf{False Positive Rate (FPR):} The ratio of all actual trajectories that are falsely identified as Sybil trajectories. This metric can be expressed as: 
		\begin{equation}
		FPR = \frac{FP}{FP + TN}
		\end{equation}
		\item \textbf{False Negative Rate (FNR):} The ratio of all Sybil trajectories that are \textit{incorrectly} identified as actual trajectories. This can be expressed as: 
		\begin{equation}
		FNR = \frac{FN}{FN + TP}
		\end{equation}
		\item \textbf{Detection Rate (DR):} The ratio of all Sybil trajectories that are \textit{correctly} identified as Sybil trajectories. The detection rate is equal to $1 - FNR$, or alternatively:
		\begin{equation}
		DR = \frac{TP}{TP+FN}
		\end{equation}
	\end{enumerate}
	
	The abovementioned metrics are used to evaluate the performance of our proposed scheme in comparison with Footprint~\cite{chang2012footprint}. We show the simulation results considering the two heuristics namely check window size and trajectory length limit explained earlier in section \ref{sec:DetectingSybilAttacks}. For each of the two heuristics and each simulation configuration, we ran the simulation 30 times and get the average of all runs.

		\begin{comment}

	\begin{figure}[!t]

		\centering
		
		\subfloat[Check window size versus false positive rate.]{
			\includegraphics[width=0.47\textwidth]{Figures_Simu/results_check_window_fpr.pdf}
		}\\
		\subfloat[Check window size versus false negative rate.]{
			\includegraphics[width=0.47\textwidth]{Figures_Simu/results_check_window_fnr.pdf}
		}\\	
		\subfloat[Check window size versus detection rate.]{
			\includegraphics[width=0.47\textwidth]{Figures_Simu/results_check_window_dr.pdf}
		}
		%\vspace{-3 mm}
		\caption{Simulation results at variable check window size.}
		\label{fig:simulation_check_window}
		%\vspace{-5mm}
	\end{figure}

		\begin{figure}[!t]
			%\setlength{\abovecaptionskip}{0.1cm} % 0.5cm as an example
			%\setlength{\belowcaptionskip}{-0.7cm} % 0.5cm as an example
			\centering
			
			\subfloat[Trajectory length limit versus false positive rate.\label{fig:a}]
			{\includegraphics[width=0.47\textwidth]{Figures_Simu/results_trajectory_length_fpr.pdf}}\\	
			\subfloat[Trajectory length limit versus false negative rate.\label{fig:b}]
			{\includegraphics[width=0.47\textwidth]{Figures_Simu/results_trajectory_length_fnr.pdf}}\\	
			\subfloat[Trajectory length limit versus detection rate.\label{fig:c}]
			{\includegraphics[width=0.47\textwidth]{Figures_Simu/results_trajectory_length_dr.pdf}}
			%\vspace{-3 mm}
			\caption{Simulation results at variable trajectory length limit.}
			\label{fig:simulation_trajectory_length}
			%\vspace{-5mm}
		\end{figure}
		
		\end{comment}
		
	\subsubsection{Impact of the check window size}
	\label{ImpactoftheCheckWindowSize}	
	In this simulation, we considered a constant trajectory length limit of 15 RSUs. Then, we changed the check window size from 2 to 50 seconds with an interval of 2 seconds. Fig.~\ref{fig:E2ECT}, Fig.~\ref{fig:E2ETP}, and Fig.~\ref{fig:E2EPDR} shows the FPR, FNR, and DR respectively at variable check window sizes for our scheme and Footprint. In both schemes, as the check window size increases, the FPR decreases. This happens because as the check window size increases, two actual trajectories are more likely to distinguish each other by having a negative similarity. Similarly, the larger the check window, the higher the FNR, since it becomes more likely that two distinct RSUs are inside the same check window and hence malicious trajectories are being falsely identified as honest. Nevertheless, the results clearly show that our scheme is better than Footprint. While the FPR remains the same as in Footprint, the FNR was decreased and the DR was increased by up to 50\% relative to Footprint, respectively. Moreover, increasing the check window size results in a lower detection rate in Footprint while in our scheme the DR is not affected by the same ratio. This is because our PoW-based approach that limits the possibility of a malicious vehicle to generate multiple trajectories simultaneously.

	\subsubsection{Impact of the trajectory length limit}
	\label{ImpactoftheTrajectoryLengthLimit}

In this simulation, we examine the effect of the trajectory length limit. We consider the check window equal to 17 seconds ("best" check window size that we obtained from previous simulation) and vary trajectory length limit from 2 to 24 with an interval of two. Fig. \ref{fig:HbHCT}, Fig.~\ref{fig:HbHTP}, and Fig.~\ref{fig:HbHPDR} plots FPR, FNR, and DR respectively as functions of the trajectory length limit. Similar to the results shown in Section \ref{ImpactoftheCheckWindowSize}, our scheme outperforms the performance of Footprint. Specifically, our proposed solution reduces the FPR by around 15\% relative to Footprint. Also, the FNR was decreased and the DR was increased by up to 40\% relative to Footprint. This is because our PoW-based approach makes it harder for a malicious vehicle to compute forged multiple trajectories simultaneously.

\subsubsection{Detection time overhead of eliminating Sybil nodes}

Finally, we consider the detection time required by the event manager to run the maximum clique algorithm and identify Sybil trajectories in our proposed scheme and Footprint~\cite{chang2012footprint} at different values of check window sizes and different trajectory length limit values. The results is shown in Fig.~\ref{checkcomptime} and Fig.~\ref{trajcomptime} respectively. Note that, the presented results represent the average of running the maximum clique algorithm $30$ times at each value. It can be seen from both Fig.~ \ref{checkcomptime}, Fig.~\ref{trajcomptime} that detection time is much less than the case in Footprint. This is because of our scheme leverage the PoW algorithm which limits the attacker's capability to create multiple successful trajectories. And hence less number of forged trajectories should be eliminated by the clique algorithm.

\begin{comment}

	\begin{figure}[!t]
		\captionsetup{justification=centering}
		\centering
		\includegraphics[width=\linewidth]{Figs/Fig1}
		
		\caption{Check window size versus detection time.}
		\vspace{0mm}
		\label{checkcomptime}
		
	\end{figure}

		\begin{figure}[!h]
			\captionsetup{justification=centering}
			\centering
			\includegraphics[width=\linewidth]{Figs/Fig2}
			
			\caption{Trajectory length limit versus detection time.}
			\vspace{0mm}
			\label{trajcomptime}
			
		\end{figure}
			\begin{figure}[!h]
				\captionsetup{justification=centering}
				\centering
				\includegraphics[width=\linewidth]{Figs/Fig3}
				
				\caption{No. of forged trajectories versus detection time.}
				\vspace{0mm}
				\label{forgecomptime}
				
			\end{figure}
			\end{comment}

	In another part of our evaluation, we choose a safe check window size ($17$ seconds) as well as a safe trajectory length limit ($15$) in our scheme and Footprint. Then, using Poisson distribution, we changed the number of forged of trajectory that a vehicle can generate. Each malicious vehicle can generate up to 40 trajectories with mean 10. Then, we run the clique algorithm and take the average of all runs. The results are shown in Fig.~\ref{forgecomptime}. Our PoW based scheme outperforms Footprint. In Footprint, as the number of Sybil trajectories a malicious vehicle can generate, the computation time to detect Sybil nodes increases and reaches several seconds. As an example, if malicious vehicles were able to generate $28$ forged trajectories, Footprint takes a round $26$ seconds. However, our scheme can detect them in milliseconds. Therefore, although Footprint can detect Sybil nodes, it suffers from high computation cost to detect Sybil trajectories making it not suitable in VANET especially in safety-related applications which require online detection of Sybil attacks in range of milliseconds which is the case in our scheme. In addition, an attacker may try to leverage the detection time burden of the clique algorithm performed by the event manger to launch denial of service attack. This attack is very hard in our scheme thanks to the PoW that we have used in our scheme.

	\subsection{Security/Privacy Analysis}
Our scheme meet the following security and privacy features,
	\begin{enumerate}
   \item \textit{Resist to RSU compromise attack:} Since RSUs are mainly responsible for issuing authorized messages for the vehicles. An attacker may try to compromise an RSU to issue forged trajectories that look like legitimate. But using threshold signatures, only one RSU cannot issue a proof of location for a specific vehicle. Alternatively, the contribution of at least $t$ RSUs are required. Therefore, our scheme mitigates any possibility of RSU compromise attack.

	  \item \textit{Resist to replay attacks:} An attacker may try to eavesdrop any of authorized messages from other honest vehicles to misuse them to create Sybil nodes. However, such an attack is prohibited in our scheme since RSUs should first verify any received message before it can issue a proof of location to that vehicle. An attacker shall know the temporary private key of the message owner to pass the message verification step which is difficult to achieve.
		
		%In our scheme, if an a malicious vehicle try to eavesdrop any of generated authorized message (owned by a honest vehicle), he/ she would fail since according to Fig~\ref{proposed}, upon arriving at $R_{2}$, $v_{i}$ should use the $Pr^{1}_{v_{1}}$ which is the private key associated with the public key $PK^{1}_{v_{1}}$ that is included in $m_{1}$ to sign $L_{R_{1}}$. Then, once $R_{2}$ receives $L_{R_{1}}$, it verifies the ownership of message $m_{1}$ by checking whether the signature received is generated by the vehicle that really owns the corresponding private key of $PK^{1}_{v_{1}}$.  

	   \item \textit{Resist to DoS attacks:}  Although Footprint can detect Sybil nodes, vehicles are free to start creating its trajectory whenever it wants. Therefore, a Sybil node can cause a denial of service attack by overwhelming honest vehicles/RSUs by a flood of forged trajectories which may cause wrong decisions such as congestion, especially in dense areas. However, in our scheme,  using the PoW approach limits the ability of a malicious vehicle to create forged trajectories as discussed in Sec.~\ref{Selection of PoW targets}.

      \item \textit{Preserve vehicles' location privacy:} Our scheme preserve the privacy of vehicle locations since the signature contained in the message is anonymous and signed by a certain threshold number of RSUs which makes an attacker unable to decide which RSU signed a particular message. Thus, no information about the location can be inferred by having an RSU signature. Also, the authorized messages issued by RSUs contain temporary location tags that change over time, therefore, if an attacker tried to memorize such location tags, this would not help in future as the RSUs changes their location tags over time.

    \item \textit{No need for a predefined design for RSUs:}
  Due to design issues in the deployment of VANETs, there may be a coverage overlap between two neighbouring RSUs transmission power. This overlap can be a severe problem, for example, in~\cite{chang2012footprint}, when a vehicle requests proof of location from an RSU, it is possible that multiple RSUs may simultaneously receive that request and issue a proof of location for this vehicle (especially in a dense deployment of RSUs). Therefore, the vehicle can obtain multiple trajectories from different RSUs that can be misused by a malicious vehicle to launch a Sybil attack. However, our scheme resists against such problem since the PoW algorithm limits a malicious vehicle from obtaining valid multiple proofs of locations since it should pass the PoW verification step for each received puzzle. Therefore, our scheme does not require a specific configuration or design for RSUs especially in the early deployment of VANETs.
    
	\end{enumerate}

	\begin{comment}
	
	\begin{table}[!t]
		\centering
		\begin{tabular}{llll}\toprule
			&   RobSAD~\cite{chen2009robust}  &  Footprint~\cite{chang2012footprint} & Proofprint \\ \midrule
			Reply attacks  & \xmark           &\cmark & \cmark \\
			
			Compromisation attack  &\xmark  &\xmark & \cmark   \\
			Heterogeneous design &\cmark  &\xmark & \cmark   \\
			\bottomrule
		\end{tabular}
		\caption{Security comparison.}
		\label{cost1}
	\end{table} 
	
	\end{comment}
	
	%\subsection{Computation and Communication Analysis}
	
	%In this section, we evaluate the computation and communication overhead of our proposed scheme.

	%\subsubsection{Computation complexity of creating trajectories}
	
	%In this section, we discuss the computation complexity of creating trajectories in terms of time to issue trajectories and the communication overhead in terms of the size of exchanged messages between the vehicles and RSUs. 
	
	\begin{comment}

	For example, as per Fig.~\ref{proposed}, considering a threshold $t=3$, $R_{3}$ should do the following:
	\begin{enumerate}
		\item  Verify $3$ shares generated by previous RSUs (i.e, $R_1$ and $R_3$) namely $\sigma_{R_{1}}(m_{1})$, $\sigma_{R_{2}}(m_{1})$ and $\sigma_{R_{2}}(m_{2})$ respectively.
		
		\item Compute $3$ signature shares namely $\sigma_{R_{3}}(m_{1})$, $\sigma_{R_{3}}(m_{2})$ and $\sigma_{R_{3}}(m_{3})$.
		
\end{enumerate}
\end{comment}
	\begin{table*}[!h]
		\caption{Overall computation and communication overhead comparison.}
		\label{communication cost}
		\centering
		\begin{threeparttable}

			\begin{tabular}{l|l|l}
				
				\hline \hline
				& Footprint~\cite{chang2012footprint} & Our scheme\\\hline
			Computation overhead ($ms$)	&  27 \textsl{Exp}   &$(1 \cdot \textsl{Exp} + 2 \cdot \textsl{Pairing})\times t$\\\hline
			Communication overhead (\textit{Bytes})	& $24 \cdot l + 372 $ & $24\cdot  l + 20\cdot t$ \\\hline
			
			\end{tabular}

			\begin{tablenotes}
				\item[*] \textsl{Exp}: group exponentiation operation (1024bit). \textsl{Pairing}: pairing operation. $l$ is the number of RSU that a vehicle encounters.
			\end{tablenotes}
		\end{threeparttable} 
	\end{table*}

	\begin{comment}
	
	\begin{table}[!t]
		\centering
		\begin{tabular}{ll}\toprule
		Cryptographic Operation 	&   Time   \\ \midrule
		Pairing $e(P_1; P_2)$  & 3.138600 ms   \\ \hline
		Hash & 0.058359  ms   \\ \hline	
		Add &  0.000227  ms   \\ \hline	
		Mul & 0.000269   ms   \\ \hline
		EXP &  0.333714  ms   \\ \hline
		\end{tabular}
		\caption{Security comparison.}
		\label{cost1}
	\end{table} 
\end{comment}

\section{Computation and Communication Overhead}
\label{Computaion overhead}
To evaluate the communication and computation overheads of our scheme, we implemented the required cryptographic operations using Python charm cryptographic library \cite{akinyele2013charm} running
on Raspberry Pi 3 devices with 1.2 GHz Processor and 1 GB RAM.
We used supersingular elliptic curve with the asymmetric Type 3 pairing of size 160 bits (MNT159 curve) for bilinear pairing to estimate the BLS signature used by our underlying threshold signature scheme.

\subsubsection{Computation overhead}
The computation overhead measures the time required to create trajectories for vehicles. Indead, the computation burden in creating trajectories lies on the RSUs since the vehicle is just need to generate a signature whenever it encounters an RSU which is cheap operation. The signing and verification operations takes $0.39$ ms and $6.27$ ms, respectively. In our scheme, when the vehicle requests a proof of location message, the RSU needs to generate only $t$ signature shares. Also, an RSU should verify $t$ signature shares to check whether a message received from a vehicle is authentic and signed by a neighboring RSU or not. Therefore, the overall computation cost by an RSU to issue a proof of location message for a vehicle as a function of $t$ is $0.39 \cdot t + 6.27 \cdot t = 6.66\cdot t$ ms while in Footprint, is about 27 modular exponentiations, which takes roughly $9.01$ ms (i.e., $110$ vehicles/second). So, to achieve trade-off between the computation overhead and security (Resist against RSU compromise), it is important to choose suitable threshold value $t$ (for different areas in VANETs). As an example, if $t=4$, RSU needs $24$ ms to issue a proof of location for a vehicle ( i.e., $41$ vehicles/second) which is practical even in urban area settings (condense areas) where the traverse time is about $20$ seconds~\cite{eriksson2008cabernet} which is sufficiently long for issuing hundreds of proof of location messages for vehicles. Therefore, our scheme is practical in urban vehicular scenario.

%To estimate the overall computation cost by an RSU to issue authorized messages for vehicles, we used Charm cryptographic library~\cite{akinyele2013charm} to measure the execution times of the cryptographic operations used in our scheme namely the BLS signature with elliptic curves $\mathbb{F}_{3^{97}}$ to evaluate the underlying threshold signature scheme. 

\subsubsection{Communication overhead}

 The communication overhead is measured by the size of transmitted message between an RSU and a vehicle in \textit{bytes} so a vehicle can have its own trajectory. A message issued by an RSU should issue 4-byte timestamp, 20-byte location tag, a 20-byte signature share size in our scheme while a 372-byte signature in Footprint. Therefore, in our scheme, the message size in \textit{bytes} exchanged between an RSU and a vehicle can be approximated as $24\cdot l + 20\cdot t$ where $l$ is the number of RSUs that a vehicle encounters and $t$ is the threshold $t$ of underlying $(t, n)-$ threshold signature protocol. Also, the communication overhead in Footprint~\cite{chang2012footprint} is $24\cdot l + 372$ bytes. It can be noted that in both our scheme and Footprint, as the trajectory of a vehicle continues to grow, the message size is increasing which consumes more communication bandwidth. To limit the size of messages and also achieve the temporary linkable feature of location tags, an event with a short period of time should be chosen. As an example, for an event with $30$ minuets with a mean of $60$ seconds as the traverse time between RSUs, in this case, $l=30$ which is relatively small and hence the less the messages' size.

\begin{comment}
We estimated the communication overhead using the following parameter sizes:
\begin{enumerate}
	\item The signature share generated by an RSU is equal to $20$ bytes. 
	\item The size of concatenation of both location tag and time stamp is $24$ bytes where the location tag as $20$ bytes and the time stamp is $4$ bytes.
	\item The PoW puzzle solution is $32$ bytes assuming SHA-256 is used in the proof of work algorithm.
	\item Assume the vehicle uses RSA public key cryptosystem with $2048$ bit moduli (so, digital signatures will be $2048$ bits long) to sign messages sent to the RSU.
\end{enumerate}
\end{comment}

%Vehicle & $24 \times t-1 + 20 \times t-1 + K_{PoW}  +  K_{PK} + K_{\sigma}$

	\section{Related work}
	\label{related}
	The detection of Sybil attacks relies on three categories, namely, identity registration, position verification and trajectory-based approaches.
	%While Douceur~\cite{douceur2002sybil} first introduced the Sybil attack, extensive works are done to detect Sybil attack.These approaches can be classified into four main classes:Identity registration, resource testing, trust-based and location verificationbased approaches.

	Identity registration approaches aims to ensure the trustworthy of each vehicle by using public key cryptographies, certificates and digital signatures. Zhou et al.~\cite{zhou2011p2dap} proposed a privacy-preserving scheme based on certificates to detect Sybil nodes. The department of motor  vehicle (DMV) represents the certificate authority, and is responsible for providing vehicles with a pool of pseudonyms to be used to hide the vehicle's unique identity. The pseudonyms associated with each vehicle are hashed to a common value. An RSU determines whether the pseudonyms come from the same pool by calculating the hashed values of the received pseudonyms. RSUs can detect Sybil nodes and then report such suspected vehicles to DMV. To resist against RSU compromise, the paper suggests two-level hash functions with different keys (coarse-grained keys and fine-grained keys). RSU holds each valid coarse-grained key only for a short time which does not know whether the pseudonyms belong to one vehicle or not. If an RSU is compromised, the attacker only gets the coarse-grained hash key for the current time interval while DMV stores all keys and can detect Sybil nodes by two-level hashing. Although deploying trusted certificates is the most efficient approach that can completely eliminate Sybil attacks, it also violates both anonymity and location privacy of entities. Also, relying on a centralized authority to ensure each is assigned exactly one identity which becomes a bottleneck in the large-scale network such as VANETs. In~\cite{wu2010balanced}, Chen et al. proposed a group signature-based approach that can be used to enable a member in the group to authenticate himself/ herself anonymously. Meanwhile, if a particular node generates multiple signatures on the same message, the verifier can  recognize those signatures. As a result, detecting duplicated signatures signed by the same vehicles can eliminate Sybil attack. However, the malicious vehicle can launch Sybil attack, if he can generate different messages with similar meaning. Recently, Reddy et al.~\cite{reddy2017sybil} proposed a cryptographic digital signature based method to establish the trust relationship among participating entities.

 Location verification based approaches are another solution to detects Sybil nodes based on physical measurements such as Received Signal Strength Indicator (RSSI) and Time Difference of Arrival (TDoA). Bouassida et al.~\cite{bouassida2009sybil} proposed a detection mechanism utilizing localization technique based on RSSI. First, consecutive RSSI variations are checked if they fall into a reasonable period or not. Some nodes which fail the test are labeled as "suspected". Then, distinguishability degrees are calculated for each suspected node by estimating its geographical localizations. Identities with the same estimated location are judged as Sybil nodes. However, the proposed scheme was tested in a small-scale testbed where the distance between two adjacent nodes was only 10m. Moreover, the scheme assumes a predictable propagation model for position estimation that may fail to capture notorious variations of wireless channels. Recently, Jin. al~\cite{jin2014traffic} proposed a detection method based on TDoA to locate the source of a message. The authors suggest installing three or more receiving sensors on a vehicle. Then, using the arrival time of a beacon message on three time-synchronized sensors,  the TDoA between the three receiving sensors can be calculated.  If the location is different from the claimed location included in the beacon message, then the node will be considered as a Sybil node. However, extra expenses during vehicle manufacturing are needed to implement this method.
	
	%RSSI-based approaches are low-cost methods. However, to estimate locations of neighboring nodes, most of these methods assume specific propagation models and the estimation efficiency depends on the chosen model. Therefore, they are  not suitable for high mobility VANETs,

The most relevant approach to our work is using trajectories of vehicles as its identities to ensure trust between participating nodes. In~\cite{chen2009robust}, RSUs broadcasts digital signatures with a timestamp to vehicles which are under its coverage. Vehicles store the RSUs signatures which they gathered in motion. However, since the time stamp is not issued for a dedicated vehicle, a malicious vehicle may claim its presence at certain RSU by merely eavesdropping such broadcasted timestamp on a wireless channel although it may have never been there at that time. In~\cite{chang2012footprint}, Footprint has been introduced to detect Sybil attack. When a vehicle passes by an RSU, it obtains a signed message as proof of presence at this location at a particular time. A trajectory of a vehicle is a consecutive series of authorized messages collected by the vehicle as it keeps traveling. Sybil attack can be detected using the fact that the trajectories generated by an attacker are very similar. However, Footprint has some critical issues. First, in Footprint, RSUs are assumed to be fully trustworthy. However, if an RSU is compromised, it can help a malicious vehicle generate fake legal trajectories by concatenating any valid tags and timestamps. In that case, Footprint cannot detect such trajectories. Second, since any vehicle is free to start its trajectory at any time, an attacker can construct multiple trajectories while moving, causing not only Sybil attacks, but also denial of service attack because of the complexity of finding similar Sybil trajectories is very high in terms of time which is a critical concern especially in safety-related applications in VANET~\cite{ma2012design}. Third, vehicles may obtain proof of appearance from multiple RSUs simultaneously (e.g., in a dense deployment). That can be used by a malicious vehicle to launch a Sybil attack . To tackle that problem, the paper suggests configuring the transmission power of RSUs properly so that there is no coverage overlap between two neighboring RSUs. That solution is hard to achieve due to wireless communication properties.

\section{CONCLUSION}
	\label{conclusion}
Sybil attacks can cause disastrous consequences in VANETS. In this paper, we have introduced a novel approach for detecting Sybil attacks using proofs of work and location. An anonymous trajectory of a vehicle is formed by obtaining a consecutive proof of locations from multiple RSUs which it encounters.  Instead of allowing only one RSU to issue authorized messages for vehicles, at least $t$ RSUs are required for creating a proof of location message using threshold signature to mitigate the RSU compromise attack. Also, the use of proof-of-work algorithm can limit the ability of malicious vehicles to create forged trajectories. Our evaluations have demonstrated that our scheme can detect Sybil attacks with high rate and low false negative rate. Moreover, the communication and computation overhead of the exchanged packets are acceptable. %In future work, we will study the case in which vehicles collude with each other to pollute the VANETs with fake messages. 

\bibliography{references}		
\bibliographystyle{IEEEtran}
	\begin{IEEEbiography}[{\includegraphics[width=1in,height=1.25in,clip,keepaspectratio]{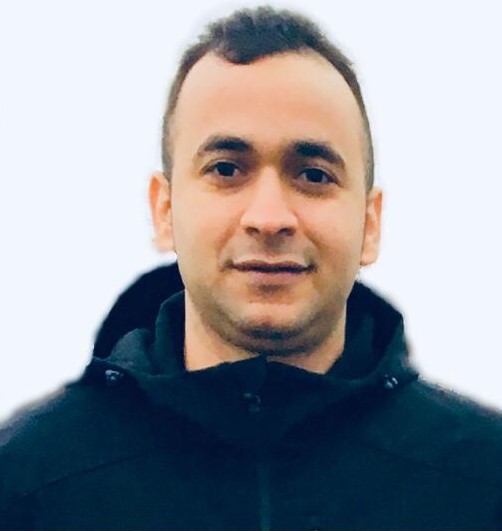}}]{\textbf{Mohamed Baza}} is currently a Graduate Research Assistant in the Department of Electrical \& Computer Engineering, Tennessee Tech. University, USA and pursuing his Ph.D. degree in the same department. He received the B.S. degree and the M.S. degree in Computer Engineering from Benha university, Egypt in 2012 and 2017, respectively. He was a recipient of the prestigious 2nd place award during his graduation in 2012. His research interests include Blockchains, cryptography and network security, smart-grid and AMI networks, and vehicular ad-hoc networks.
		
	\end{IEEEbiography}
	
	\begin{IEEEbiography}[{\includegraphics[width=1in,height=1.25in,clip,keepaspectratio]{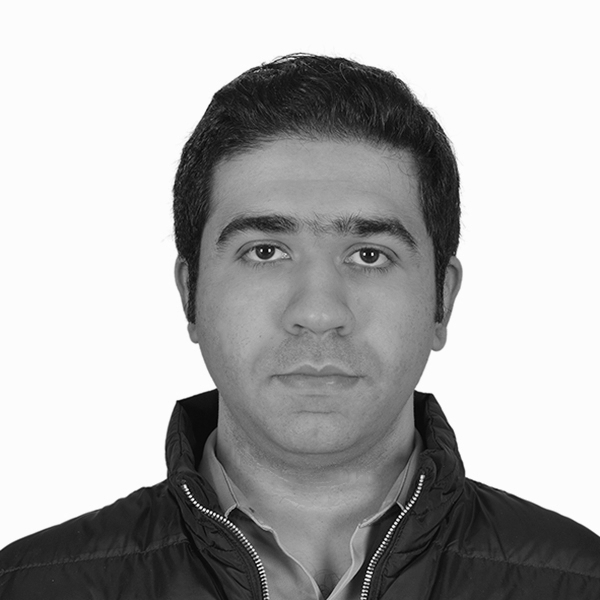}}]{\textbf{Mahmoud Nabil}} is currently a Graduate Research Assistant in the Department of Electrical \& Computer Engineering, Tennessee Tech. University, USA and pursuing his Ph.D. degree in the same department. He received the B.S. degree and the M.S. degree in Computer Engineering from Cairo University, Cairo, Egypt in 2012 and 2016, respectively.
		His research interests include machine learning, cryptography and network security, smart-grid and AMI networks, and vehicular ad-hoc networks.
	\end{IEEEbiography}

	\begin{IEEEbiography}[{\includegraphics[width=1in,height=1.25in,clip,keepaspectratio]{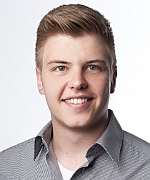}}]{\textbf{Niclas Bewermeier}} is currently working towards a Master of Science degree in Electrical and Computer Engineering at Tennessee Tech University, where he has worked as a Graduate Research and Teaching Assistant since 2016. His research focuses on privacy preservation and security threads in Vehicular Ad-Hoc Networks. He is expected to graduate in May of 2019. In 2012, Niclas began his studies at Cologne University of Applied Sciences, where he graduated in 2016 with a Bachelor of Science degree in Electrical Engineering.
	\end{IEEEbiography}
	
	\begin{IEEEbiography}[{\includegraphics[width=1.1in,height=1in,clip,keepaspectratio]{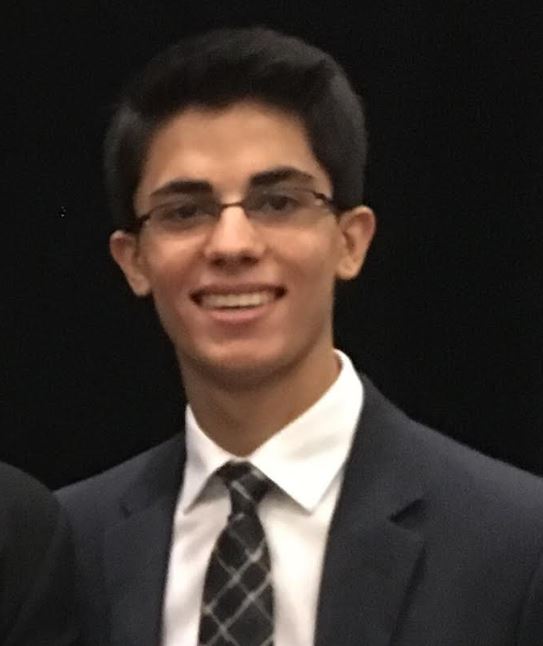}}]{\textbf{Kemal Fidan}} is currently an undergraduate at The University of Tennessee, Knoxville pursuing a degree in computer science. At UT, he works for the NSF and DOE funded research center, CURENT. His research interests include security of blockchain technology, and its various applications.
	\end{IEEEbiography}

	\begin{IEEEbiography}[{\includegraphics[width=1in,height=1.25in,clip,keepaspectratio]{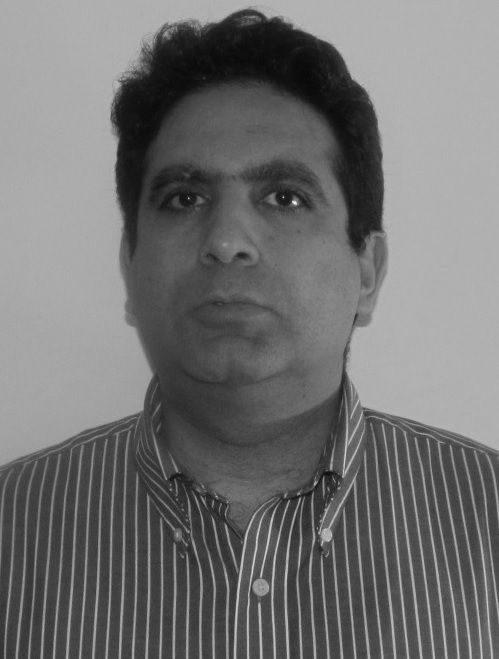}}]{\textbf{Dr. Mohamed M. E. A. Mahmoud}}
		received PhD degree from the University of Waterloo in April 2011. From May 2011 to May 2012, he worked as a postdoctoral fellow in the Broadband Communications Research group - University of Waterloo. From August 2012 to July 2013, he worked as a visiting scholar in University of Waterloo, and a postdoctoral fellow in Ryerson University. Currently, Dr Mahmoud is an associate professor in Department Electrical and Computer Engineering, Tennessee Tech University, USA. The research interests of Dr. Mahmoud include security and privacy preserving schemes for smart grid communication network, mobile ad hoc network, sensor network, and delay-tolerant network. Dr. Mahmoud has received NSERC-PDF award. He won the Best Paper Award from IEEE International Conference on Communications (ICC'09), Dresden, Germany, 2009. Dr. Mahmoud is the author for more than twenty three papers published in major IEEE conferences and journals, such as INFOCOM conference and IEEE Transactions on Vehicular Technology, Mobile Computing, and Parallel and Distributed Systems. He serves as an Associate Editor in Springer journal of peer-to-peer networking and applications. He served as a technical program committee member for several IEEE conferences and as a reviewer for several journals and conferences such as IEEE Transactions on Vehicular Technology, IEEE Transactions on Parallel and Distributed Systems, and the journal of Peer-to-Peer Networking.
	\end{IEEEbiography}

	\begin{IEEEbiography}[{\includegraphics[width=1in,height=1.25in,clip,keepaspectratio]{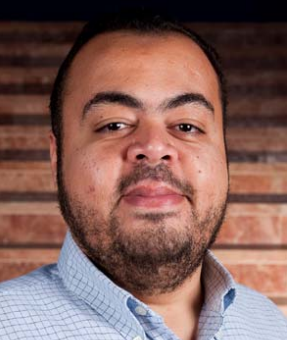}}]{\textbf{Mohamed Abdallah}} was born in Giza, Egypt. He received the B.Sc. degree with honors from Cairo University, Giza, Egypt, in 1996, and the M.Sc. and Ph.D. degrees in electrical engineering from University of Maryland at College Park, College Park, MD, USA, in 2001 and 2006, respectively. He joined Cairo University in 2006 where he holds the position of Associate Professor in the Electronics and Electrical Communication Department. He is currently an Associate Research Scientist at Texas A\&M University at Qatar, Doha, Qatar. His current research interests include the design and performance of physical layer algorithms for cognitive networks, cellular heterogeneous networks, sensor networks, smart grids, visible light and free-space optical communication systems and reconfigurable smart antenna systems.
	\end{IEEEbiography}

\end{document}